\documentclass[aps,pra,twocolumn,groupedaddress,nofootinbib,notitlepage,showpacs,floatfix]{revtex4-1}
\usepackage{graphicx,subfigure}
\usepackage{amsmath}
\usepackage{amsthm}
\usepackage{amssymb}
\usepackage[usenames]{color}
\usepackage{hyperref}
\usepackage{array}
\newcolumntype{L}{>{\centering\arraybackslash}m{3.5cm}}
\newcolumntype{R}{>{\arraybackslash}m{3.5cm}}

\hypersetup{
    colorlinks=true,       
    linkcolor=cyan,          
    citecolor=magenta,        
    filecolor=magenta,      
    urlcolor=cyan,           
    runcolor=cyan
}

\usepackage[dvipsnames]{xcolor}

\newcommand{\bes} {\begin{subequations}}
\newcommand{\ees} {\end{subequations}}
\newcommand{\bea} {\begin{eqnarray}}
\newcommand{\eea} {\end{eqnarray}}
\newcommand{\beq}{\begin{equation}}
\newcommand{\eeq}{\end{equation}}
\newcommand{\mcal}[1]{\mathcal{#1}}
\newcommand{\mc}[1]{\mathcal{#1}}
\newcommand{\D}{\Delta}
\DeclareMathOperator*{\Tr }{Tr}

\newcommand{\HTIM}{H_{\text{TIM}}}

\newtheorem{theorem}{Theorem}[section]

\newtheorem{definition}{Definition}

\begin{document}

\title{Prospects for Quantum Enhancement with Diabatic Quantum Annealing}
\author{E.J. Crosson}
\affiliation{Center for Quantum Information and Control (CQuIC), Department of Physics and Astronomy,
University of New Mexico, Albuquerque, NM 87131, USA}
\author{D.A. Lidar}
\affiliation{Departments of Electrical \& Computer Engineering, Chemistry, Physics \& Astronomy\\
Center for Quantum Information Science \& Technology (CQIST)\\
University of Southern California, Los Angeles, California 90089, USA}

\begin{abstract}
We assess the prospects for algorithms within the general framework of quantum annealing (QA) to achieve a quantum speedup relative to classical state of the art methods in combinatorial optimization and related sampling tasks.  We argue for continued exploration and interest in the QA framework on the basis that improved coherence times and control capabilities will enable the near-term exploration of several heuristic quantum optimization algorithms that have been introduced in the literature.  These continuous-time Hamiltonian computation algorithms rely on control protocols that are more advanced than those in traditional ground-state QA, while still being considerably simpler than those used in gate-model implementations.  The inclusion of coherent diabatic transitions to excited states results in a generalization called \emph{diabatic quantum annealing} (DQA), which we argue for as the most promising route to quantum enhancement within this framework.  Other promising variants of traditional QA include reverse annealing and continuous-time quantum walks, as well as analog analogues of parameterized quantum circuit ansatzes for machine learning.  Most of these algorithms have no known (or likely to be discovered) efficient classical simulations, and in many cases have promising (but limited) early signs for the possibility of quantum speedups, making them worthy of further investigation with quantum hardware in the intermediate-scale regime.  We argue that all of these protocols can be explored in a state-of-the-art manner by embracing the full range of novel out-of-equilibrium quantum dynamics generated by time-dependent effective transverse-field Ising Hamiltonians that can be natively implemented by, e.g., inductively-coupled flux qubits, both existing and projected at application scale.
\end{abstract}

\maketitle


\section{Introduction}

Quantum annealing (QA) is a heuristic algorithm typically employed for solving optimization problems formulated in terms of finding ground states of classical Ising spin Hamiltonians~\cite{kadowaki_quantum_1998}. QA started as a theoretical combinatorial optimization method (see~\cite{Albash-Lidar:RMP} for a historical survey along with the original references), while in its more recent incarnation it is also understood as a heuristic optimization method implemented in physical hardware~\cite{Dwave}. 

The idea that QA can be a powerful optimization heuristic is driven by the expectation that  
quantum fluctuations can sometimes be more efficient than classical fluctuations---such as thermal fluctuations used in classical simulated annealing~\cite{kirkpatrick_optimization_1983}---in searching for configurations of variables that minimize a cost function. As such, a physical implementation in quantum hardware is of particular interest, since it might lead to quantum speedups over algorithms running on classical hardware. Following pioneering QA experiments in naturally occurring disordered magnets~\cite{Brooke1999,brooke_tunable_2001}, theoretical proposals~\cite{2002quant.ph.11152K,Kaminsky-Lloyd} inspired the construction of commercial, non-universal quantum annealing processors by D-Wave Systems Inc.~\cite{harris_flux_qubit_2010,Harris:2010kx,Berkley:2010zr,Dwave,Berkley:2013bf,Bunyk:2014hb,DWave-entanglement,DWave-16q}. These are the largest programmable quantum information processors constructed to date, featuring thousands of superconducting flux qubits, and have driven much of the explosion of interest in QA in the last several years. Alternative superconducting approaches to QA with more coherent flux qubits are also being pursued, but have not yet approached similar scales~\cite{Yan:2016aa,Weber:2017aa,Quintana:2017aa,Novikov:2018aa,khezri2020annealpath}, as is true for Rydberg atoms, which in theory feature high programmability and long-range connectivity~\cite{Glaetzle:2017aa,qiu2020programmable}. Recently, the D-Wave processors were used as quantum simulators~\cite{Harris:2018aa,King:2018aa,Gardas:2018aa,King:2019aa,Weinberg:2020aa,b2020probing}, thus joining gate-based approaches such as ion traps~\cite{Islam:2013mi,Smith:2016aa,Zhang:2017aa}, quantum gas microscopes~\cite{Simon:2011aa,Boll:2016aa}, Rydberg atoms~\cite{Weimer:2010aa,Nguyen:2017aa,Bernien:2017aa}, and transmon-based superconducting qubits~\cite{Barends:2015kl,Barends:2015aa,OMalley:2016aa,Wang:2020aa}. This excursion to quantum simulation  demonstrates that quantum annealing has started to transcend its original scope of heuristic optimization, connecting the field to the historical origins of quantum computation~\cite{Feynman1,Lloyd:96}.

However, the lack of empirical evidence---despite numerous attempts---of an unequivocal quantum speedup in the primary application domain of heuristic optimization ~\cite{q108,speedup,Venturelli:2014nx,DW2000Q,Hen:2015rt,Katzgraber:2015gf,Rieffel:2015aa,McGeoch,PhysRevX.6.031015,2016arXiv160401746M,King:2015cs,Vinci:2016tg,Job:2017aa,Albash:2017aa,Mandra:2017ab,Adame:2018aa,Das:2019aa,sahai2019estimating}, or a clear quantum advantage in machine learning contexts relying on sampling from a distribution that includes excited states~\cite{Smelyanskiy:2012aa,2012arXiv1204.2821S,Dorband:2015aa,Lokhov:2016aa,OMalley:2017aa,Levit:2017aa,Mott:2017aa,Li:comp-bio-2017,Benedetti:2018aa,Neukart:2018aa,Li:2019aa,Zlokapa:2019aa,Zlokapa:2019ab,Cormier:2019aa,Zlokapa:2019ab,vinci2019path,Willsch:2020aa}, 
justifies a re-evaluation of the prospects of a quantum enhancement using the traditional QA approach. Here we undertake such a re-evaluation from a theoretical perspective, and offer what we believe are the most promising directions forward for the QA field. 
We begin, in Sec.~\ref{sec:CTIME-DQA}, with a definition of the corresponding computational models, and then describe the structure of the rest of this Perspective.

\section{Computational models}
\label{sec:CTIME-DQA}

\subsection{Transverse-Field Hamiltonian Interpolation (TF-HI)}

We consider general time-dependent Hamiltonians $H(t)$, but focus primarily on transverse-field 
Hamiltonian interpolation (TF-HI) models:
\begin{align}
    \label{eq:h}
    H_{\text{TFHI}}(t) &= A(t) H_X + B(t) H_Z  \ , \quad    H_X = -\sum_{i} X_i ,
\end{align} 
where $X_i$ denotes the Pauli $\sigma^x$ matrix acting on qubit number $i$, and $H_Z$ is a Hamiltonian that is diagonal in the computational basis, i.e., the basis of eigenstates of tensor products of $Z_i$ (the Pauli-$z$ matrix).

Standard, ``forward" QA is the case when $A(t)$ is monotonically decreasing to zero while $B(t)$ is monotonically increasing from zero in the interval $[0,T]$, and $H_Z$ is the problem Hamiltonian, i.e., the Hamiltonian whose ground state encodes the solution to the computational (optimization) problem. We will also discuss reverse QA, where $A$ first increases and then decreases, and where $H_Z$ is the sum of a problem Hamiltonian and another diagonal Hamiltonian, and $B(t)$ is correspondingly replaced by two separate functions of time.  The distinction between these two protocols will play an important role in our discussion of the power of the TH-HI model, below. 

A case of special interest is the transverse-field Ising model (TIM):
\bes
\begin{align}
    \label{eq:HTIM}
    \HTIM(t) &= A(t) H_X + B(t) H_Z  \\
    H_Z  &= -\sum_{i\in\mcal{V}} h_i Z_i - \sum_{(i,j)\in\mcal{E}} J_{ij} Z_i Z_j ,
    \label{eq:Hz}
\end{align} 
\ees
for $n$ qubits occupying the vertices $\mcal{V}$ of a graph $\mcal{G}=\{\mcal{V},\mcal{E}\}$.
The parameters $h_i$ and $J_{ij}$ are controllable longitudinal local field and coupling constants, respectively. When $J_{ij}>0$ ($J_{ij}<0$) the coupling is ferromagnetic (antiferromagnetic).

The time-dependence of the transverse and longitudinal coupling strengths $A(t)$ and $B(t)$ defines the annealing schedule.  We consider the broadest class of schedules compatible with realistic experimental constraints, which impose limits on the magnitudes of both $A(t)$ and $B(t)$, and their derivatives. Similarly, experimental constraints limit the precision with which the longitudinal fields $h_i$ and couplings $J_{ij}$ can be implemented. Analog control errors which modify the implemented values of the $h_i$ and $J_{ij}$ parameters are an important source of error that must be addressed~\cite{Young:2013fk,Zhu:2015pd,Albash:2019ab,Pearson:2019aa}.

Note that $\HTIM(t)$ is simpler from a control perspective than the Hamiltonians that are usually designed for gate-based quantum optimization (e.g., for QAOA~\cite{farhi2014quantum}), yet, as we shall argue, the evidence to date indicates it is at least as powerful. For this reason we do not consider more complicated variants here, such as those allowing individual control of the transverse local field terms~\cite{Adame:2018aa,Susa:2018ab}. 
However, a potential disadvantage in the lack of individual controllability of each term of $\HTIM(t)$ is a lack of modularity.  As a result, calibration of a device implementing evolution generated by $\HTIM(t)$ is considerably more complicated and time-consuming than a gate-model device, where in principle (neglecting cross-talk) each single-qubit and two-qubit gate can be calibrated independently. It is also more difficult to develop quantum error correction methods since the dynamics needs to be analyzed as a whole instead of in terms of modular components, and indeed, to date there does not exist a proof of fault tolerance for the computational model associated with $\HTIM(t)$.  Therefore we argue for investigating and implementing this computational model in the spirit of intermediate-scale algorithmic exploration, with the understanding that useful discoveries could eventually be ported to a fault-tolerant gate-model using quantum simulation.  It is also important to recognize that the high overheads associated with existing gate-model fault-tolerance schemes may preclude useful enhancement at the (finite) scale of applications (see Sec.~\ref{sec:noise} for additional discussion), and this further motivates the consideration of alternative computational models and error suppression schemes.  

We note that as written, the Hamiltonian $\HTIM(t)$ is an idealization that neglects ``leakage" states that are inevitably present in the physical realization of qubits such a flux qubits~\cite{Mooij:99} or transmons~\cite{transmon-invention}. Leakage is an important error source that needs to be carefully addressed, though we note that such higher energy states can also be beneficial and are routinely employed, e.g., to implement Raman transitions in trapped ion qubits~\cite{Islam:2013mi,Smith:2016aa,Zhang:2017aa}.

For a closed system the dynamics generated by continuous-time evolution with a general time-dependent Hamiltonian $H(t)$ corresponds to the unitary evolution 
\beq
U(T) = \mathcal{T}_+ \int_0^{T} dt\; e^{-i t H(t)} ,
\label{eq:U}
\eeq 
where $\mathcal{T}_+$ denotes forward time-ordering and $T$ is the total evolution time (see also Ref.~\cite{kendon2020quantum} for a survey of quantum computing using continuous-time evolution).
When specializing to $\HTIM(t)$, i.e., for
\beq
U_{\text{TIM}}(T) = \mathcal{T}_+ \int_0^{T} dt\; e^{-i t \HTIM(t)}  \quad \text{(CTIME)} ,
\eeq
we refer to this unitary evolution as Continuous-Time Transverse Ising Model Evolution (CTIME). 

We are interested not only in the adiabatic limit, which is the usual limit in QA, but quite explicitly also in the diabatic setting, where
the unitary evolution $U(T)$ does not follow the instantaneous energy eigenstates of $H(t)$, and diabatic transitions to and from low-energy excited states are permitted. We refer to the computational model that is implemented in this case as ``diabatic quantum annealing".\footnote{As far as we know the first time the term ``diabatic quantum annealing" appeared in the literature was in Ref.~\cite{Muthukrishnan:2015ff}, though it would be more appropriate to attribute the first appearance to Ref.~\cite{katsuda2013nonadiabatic} which called the model ``nonadiabatic quantum annealing".} 

\subsection{Diabatic Quantum Annealing (DQA)}

To more precisely define DQA, let us first recall the definition of adiabatic quantum computing (AQC). For a rigorous definition see Ref.~\cite{Albash-Lidar:RMP}; for our purposes it suffices to define AQC as the computational model in which \emph{the system at all times remains in an instantaneous eigenstate (or degenerate eigenspace) of $H(t)$}. The adiabatic theorem for closed systems provides a sufficient condition for this to hold, informally stated as $T \geq \frac{1}{\epsilon} \frac{\|\dot{H}\|}{\Delta^2}$, where $\Delta$ is the energy gap between the eigenvalue of the instantaneous eigenstate and the nearest distinct energy eigenvalue, $\epsilon$ is the distance between the latter eigenstate at $T$ and the actual state reached under $U(T)$, and the dot denotes differentiation with respect to the dimensionless time $s=t/T$ (a rigorous statement is given in Theorem~\ref{th:JRS} below).

Whereas AQC is universal in the closed system setting~\cite{aharonov_adiabatic_2007,MLM:06,Lloyd:2016}, QA is concerned with adiabatically solving optimization problems formulated in terms of classical target Hamiltonians, and is traditionally defined directly in terms of $\HTIM(t)$~\cite{kadowaki_quantum_1998}. We shall use the same universal \textit{vs} optimization distinction between AQC and QA also in the open system setting, which we discuss later.
This aligns with 
the historical origins of QA as a theoretical heuristic combinatorial optimization method~\cite{Apolloni:1989fj,Apolloni:88,finnila_quantum_1994,kadowaki_quantum_1998,Santoro} (see Refs.~\cite{RevModPhys.80.1061,Hauke:2019aa} for reviews), as well as with its more recent incarnation as an optimization method implemented in physical hardware, e.g., by D-Wave~\cite{Dwave}. 

\begin{figure}
\includegraphics[width=.8\columnwidth]{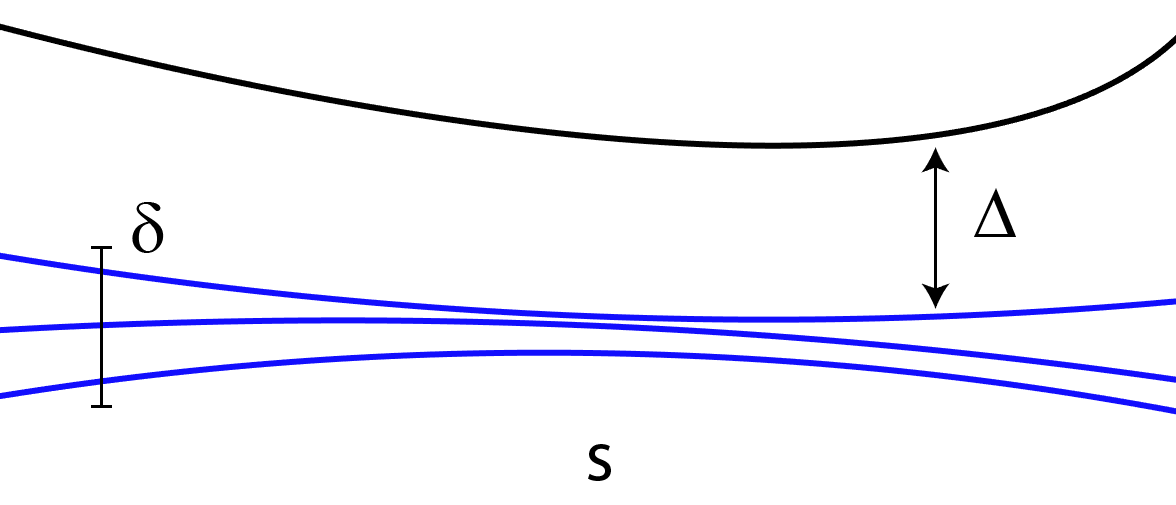}
\caption{An illustration of the lowest 4 energy levels of a parameterized Hamiltonian $H(s)$.  A diabatic quantum computation could remain contained in the subspace $\mathcal{C}$ defined by the lowest 3 energy levels, highlighted in blue.  The subspace has an instantaneous energy width $\delta(s)$ and is separated from the rest of the spectrum by an instantaneous gap $\Delta(s)$. }
\label{fig:levels}
\end{figure}

With these notions in place, we are ready to define diabatic quantum computing (DQC) and diabatic quantum annealing (DQA). Namely, when we use the adjective ``diabatic", we relax the condition that the system must at all times remain in a single instantaneous eigenstate of $H(t)$. Instead, DQC and DQA are computational models (universal and for optimization, respectively) in which \emph{the system at all times remains in a subspace spanned by eigenstates of $H(t)$ which belong to some narrow, contiguous energy band (the intersection of the Hamiltonian spectrum with an interval)} of width $\delta$, illustrated in Fig.~\ref{fig:levels}.  In general this energy band contains multiple distinct energy levels, and diabatic transitions between these are allowed, while excitations out of the band are suppressed.  Although  these definitions are stated in a general form that applies to any part of the energy spectrum, the instantaneous computational eigenstate in AQC or QA is almost always taken to be the ground state of $H(t)$, and the energy band in DQC is correspondingly taken to be the low-energy subspace of $H(t)$. We subsequently specialize to this case throughout most of the following.

Note that in DQC the final state reached under $U(T)$ need not be close to the ground state of $H(t)$, and likewise in DQA the final state reached under $U_{\text{TIM}}(T)$ need not be close to the ground state of $\HTIM(t)$. Note further that DQA dynamics are generically non-local for all $T > 0$,\footnote{$U_{\text{TIM}}$ cannot be written as a tensor product of local unitary gates due to the noncommutativity of the terms in Eq.~\eqref{eq:HTIM}.} and the  standard method for digitizing this non-local unitary on a gate-model device requires $\Omega(n T)$ gates to accurately approximate $U_{\text{TIM}}$~\cite{haah2018quantum}.\footnote{Recall that $f(x)=\Omega(g(x))$ means that $f$ is bounded below by $g$ in the large $x$ limit. I.e., informally $\Omega$ means ``at least".}  This means that DQA hardware can allow for the study of intermediate-scale quantum algorithms that are distinct from, and complimentary to, those being implemented on gate-model devices. 

The emphasis on the dynamics remaining confined to a low energy subspace is the key difference between the diabatic model and the gate-model. Such a restriction is not imposed as part of the definition of the gate model, and in principle the latter allows for excitations of arbitrarily high energy, though in practice there is of course a limit set by finite energy resources, finite bandwidth, etc.\footnote{It is also worth commenting on the distinction between continuous (e.g., CTIME) and discrete time (gate-model) evolutions. This distinction is somewhat artificial since in practice evolution is also continuous in the gate-model, when it is viewed, as it should, as generated by differentiable Hamiltonians.}

It is natural to ask what replaces the adiabatic condition in the diabatic setting. Interestingly, the general form of the adiabatic theorem as stated, e.g, in Ref.~\cite{Jansen:07} is already sufficient; this theorem gives conditions for preserving any low energy subspace (not just the ground subspace) separated by a gap from the rest of the spectrum of $H(t)$. We present this theorem here since it allows us to give precise sufficient conditions for DQC and DQA.

Consider a closed quantum system evolving for a total time $T$ subject to the Hamiltonian $H(t)$ acting on the Hilbert space $\mc{H}$. Defining the rescaled (dimensionless) time $s=t/T$, the evolution is governed by the unitary operator $U(s)$ [the same as Eq.~\eqref{eq:U}] which is the solution of
\begin{equation}
    U'(s) = -iT H(s) U(s), \quad U(0) = I, \quad s\in[0,1] ,
    \label{eq:exact}
\end{equation}
where the prime denotes differentiation with respect to $s$.\footnote{We use both $s$ and $t$ in this paper, as convenient, with $s(t)$ always denoting dimensionless time. Note that $s(t)$ need not always be $t/T$, and can in general be an arbitrary function of time.}
Let $P(s)$ be a finite-rank projection on the low-energy subspace of $H(s)$, i.e., the subspace $\mc{C}$ spanned by the eigenvectors of $H(s)$ with the lowest $d$ eigenvalues.

\begin{theorem}[Theorem 3 of Ref.~\cite{Jansen:07}]
\label{th:JRS}
Suppose that the spectrum of $H(s)$ restricted to $P(s)$ consists of $d(s)$ eigenvalues (each possibly degenerate, crossing permitted) separated by a gap $\D(s)$ from the rest of the spectrum of $H(s)$, and $H$, $H'$, and $H''$ are bounded operators. Let $P_{T}(s)\equiv U(s) P(0) U^\dag(s)$ and $X|_{a,b} \equiv X(a) + X(b)$. Then:\footnote{This theorem can be generalized so that the assumption of boundedness of $H$, $H'$, and $H''$ can be avoided~\cite{ML:tbp}. We also note that there exist tighter bounds than implied by Theorem~\ref{th:JRS} under different assumptions; see Ref.~\cite{Albash-Lidar:RMP} for a survey of different forms of the adiabatic theorem.}
\bes
\label{eq:JRS-AT}
\begin{align}
\label{eq:JRS-AT-a}
& \|P_{T}(s)-P(s)\| <\frac{\xi(s)}{T} \\ 
\label{eq:JRS-AT-b}
& \xi(s) = \left.\frac{d\|H'\|}{\D^2}\right|_{0,s} + \int_0^{s}\left(\frac{d\|H''\|}{\D^2} + 7d\sqrt{d}\frac{\|H'\|^2}{\D^3} \right)ds' .
\end{align}
\ees
\end{theorem}
This theorem means that as long as $\mc{C}$ is separated by a gap $\D$ from the higher excited states (see Fig.~\ref{fig:levels}), an evolution that starts in $\mc{C}$ will remain in this subspace, up to an error bounded by ${\xi(s)}/{T}$, i.e., an error that can be made smaller by increasing $T$. Abrupt changes in $H(s)$ and a large value of $d$ both contribute to a larger error. 

An important quantity is the width $\delta$ of the energy band inside $\mc{C}$, i.e., the difference between the top and lowest eigenvalues of $P(s)H(s)P(s)$. The key point is that, as we shall review below, arguments about the possibility of efficient classical simulation of QA in the closed system case depend strongly on the assumption that $\delta(s)=0$ $\forall s$, i.e., the system remains at all times in the ground subspace of $\HTIM$. However, Theorem~\ref{th:JRS} shows that the same conditions which are typically assumed for such ground state evolution already ensure a more general scenario that captures DQC and DQA, which we can now define as the computational models for which $\delta(s)>0$ $\forall s$. This condition sets up the distinction between AQC and DQC and between QA and DQA. Moreover, in the gate model $P(s)$ is the identity operator (there is no projection to a low-energy subspace) and hence $\delta(s)$ scales extensively with $n$. To clearly separate DQC and DQA from the gate model we impose the additional condition that $\delta(s)=\mc{O}(1)$ $\forall s$.\footnote{A gate-model device that executes layers of gates in parallel will generically have $\delta = \Omega(n)$, so $\delta = \mc{O}(1)$ is already enough to distinguish DQA from the gate-model.} 

Theorem~\ref{th:JRS} also implies that, in principle, imposing conditions that correspond to DQA are not any harder than those corresponding to QA, since the degree of leakage out of the ground subspace or out of $\mc{C}$ is controlled by the same quantities in both cases, but these quantities depend on the choice of $\mc{C}$. To illustrate this point, consider that while for the same Hamiltonian $\HTIM(s)$ it is by definition true that $d(s)$ is larger for DQA than for QA, it may well be the case that it is possible to choose the subspace $\mc{C}$ such that the gap $\D(s)$ is always larger for DQA than for QA (this is indeed the case in the ``glued-trees" problem~\cite{Somma:2012kx}, discussed below). These considerations show that DQA may be the easier computational model to implement, at least in terms of the sufficient condition for preserving the subspace $\mc{C}$ given by Theorem~\ref{th:JRS}, and the fact that this may not return the ground state of $\HTIM(s=1)$ may be offset by the fact that efficient classical simulation is thwarted as well.

\subsection{Structure of the remainder of this Perspective}
Section~\ref{sec:stoq} is concerned with optimization using transverse-field Hamiltonian interpolation. We draw a crucial distinction not only between QA and DQA, but also between coherent and weakly-decoherent evolution, and evaluate the prospects for enhancement according to existing evidence in the literature that supports the possibility of a speedup, and according to the expected overhead of classical simulation. 

In Sec.~\ref{sec:alg2} we survey several heuristic optimization algorithms defined in the literature by protocols involving initial state preparation and choices of $\{A(t),B(t)\}$, and again evaluate the prospects for enhancement. A substantial computational overhead for classical simulations is a desirable condition for quantum enhancement, and we are often interested in examples for which classical simulations are intractable (i.e., quantum processes for which state-of-the-art simulation algorithms require time exponential in $n,T$).\footnote{Note that this is not a strict requirement, since polynomial quantum speedups are the most that can be reasonably expected for NP-hard optimizaton problems.}  Arguments for the intractability of classical simulations can be supported by formal complexity-theoretic evidence or by empirical evidence (i.e., determining whether the quantum processes can currently be simulated with existing state-of-the-art classical algorithms and machines). Regarding empirical evidence, there is a general consensus that quantum dynamics are classically intractable beyond a few special cases (e.g., stabilizer circuits~\cite{Aaronson:2004aa}, matchgate circuits~\cite{jozsa-2008}, circuits defined over planar graphs with a restricted number of non-nearest neighbor gates~\cite{Geraci:2010gf}, etc.), and so we argue that diligent exclusion of these cases by consideration of all existing simulation methods allows one to exclude efficient classical simulations with high confidence. 

Whereas Secs.~\ref{sec:stoq} and~\ref{sec:alg2} focus on the case of stoquastic Hamiltonians, Sec.~\ref{sec:nonstoq} comments on the role of non-stoquastic Hamiltonians in classical intractability. We point out that non-stoquasticity is desirable but not essential for quantum enhancement. 

In Sec.~\ref{sec:formal} we critically examine the application of formal complexity theoretic arguments to real devices, and also discuss the important distinction between enhancement and classical intractability in sampling.

In Sec.~\ref{sec:noise} we consider the issue of noise as it affects the prospects for enhancement with both QA hardware and NISQ-era gate-model devices.   The phenomenon of $J$-chaos  (which is closely analogous to coherent gate errors) places limitations on the size of QA devices that can be expected to accurately solve optimization instances.   The precise relation between errors in the cost Hamiltonian and errors in the output distribution is complicated in general, but if we assume the output distribution is approximately thermal then intrinsic control errors on the order of a few $\%$ will make the fidelity with the intended output distribution nearly zero above several hundred qubits~\cite{Albash:2019ab}.  A similar problem occurs under the assumption that an output distribution remains thermal at a constant temperature as the system size increases~\cite{Albash:2017ab}.   Assuming control errors cannot be further reduced this implies that some amount of error suppression~\cite{Pearson:2019aa} or fault-tolerance would be required to reach the application scale ($n \sim 10000$ logical qubits).   
Noise limits the size of NISQ-era gate-model computations in a very similar way~\cite{Zlokapa:2020}, and therefore no NISQ algorithm for heuristic optimization can provide a quantum enhancement at the application scale without first being implemented on future hardware that includes some level of error correction or fault-tolerance.   A problem with this approach is that the time and space overhead used in fault-tolerant quantum computation is so large on practical scales (even though it is asymptotically polylogarithmic) that it can can erode a polynomial (e.g., quadratic, Grover-like) advantage~\cite{campbell2019applyingquantum,s2020compilation}.   This implies that, whether one considers QA or gate-model hardware, finding any quantum algorithm that enhances optimization at the application scale will require (1) error suppression methods that greatly reduce the overhead from that needed for fault-tolerance, and which may not decrease noise arbitrarily in the manner of a threshold theorem but which can reduce control errors enough to become useful at $n \sim 10000$, and/or (2) the discovery of heuristic quantum optimization algorithms that give stronger speedups (super-quadratic at least), which could be useful even in light of current estimates of fault-tolerance overhead.   

With these considerations in mind we conclude in Sec.~\ref{sec:summary} and advocate for the continued development of QA hardware for the primary purpose of algorithmic exploration in the intermediate-scale regime (hundreds of qubits), and for the secondary purpose of developing Hamiltonian error suppression methods that increase the scale of optimization problems which can be solved, without requiring the full overhead of fault-tolerance.

\section{Optimization using Transverse-Field Hamiltonian Interpolation}
\label{sec:stoq}

\begin{figure*}[t]
\includegraphics[width=.9\textwidth]{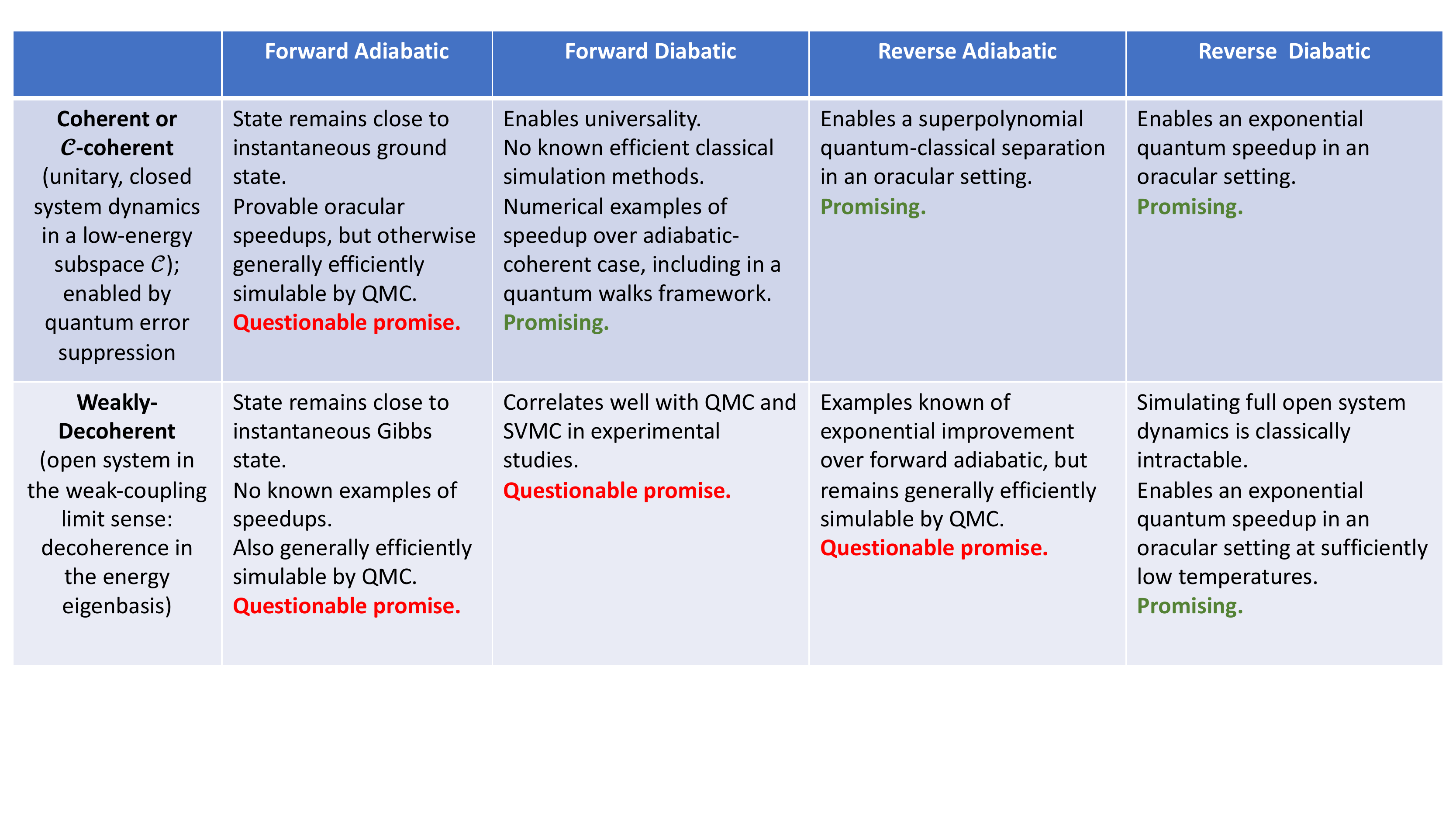}
\vspace{-1cm}
\caption{The (a)diabatic/(de)coherent division for \emph{stoquastic} forward and reverse annealing. In the forward annealing case (second and third columns), the coherent/$\mc{C}$-coherent-diabatic case is the most promising since it enables universality and does not admit efficient classical simulation methods. The other three forward annealing cases are of questionable promise due to results showing close correlation with efficient simulation with QMC. In the reverse annealing case (fourth and fifth columns), there are strong theoretical results supporting a super-polynomial quantum advantage in the oracular setting, and even the weakly-decoherent diabatic case is promising, both for quantum simulation and in the oracular setting. It is crucial to note that the coherent/$\mc{C}$-coherent case requires Hamiltonian error suppression in order to enforce coherence in the low-energy subspace $\mc{C}$.}
\label{fig:4-cases}
\end{figure*}

The most well studied algorithms based on Eq.~\eqref{eq:h} 
are defined by starting with a monotonically decreasing function $A(t)$ and monotonically increasing function $B(t)$, with the boundary conditions $A(0) > 0$, $B(0) = 0$ and $A(T) = 0$, $B(T) > 0$ (we discuss other schedules below). In this setting $T$ is called the anneal time, and TF-HI is also known as ``forward" QA. A very important property of Eq.~\eqref{eq:h} with far-reaching consequences is that the Hamiltonian is stoquastic, i.e., its matrix elements are all real and nonpositive~\cite{Bravyi:QIC08}. We defer our discussion of nonstoquastic Hamiltonians to Sec.~\ref{sec:nonstoq}.

\subsection{Adiabatic \textit{vs} diabatic}
In accordance with our previous discussion, we classify TF-HI as adiabatic if the system is initialized in the ground state of $H(0)$ and the timescale $T$ is sufficiently long to keep the state of the system close to the instantaneous ground state of $H(t)$ for all $t\in[0,T]$ [i.e., the bound in Eq.~\eqref{eq:JRS-AT-a} is smaller than some constant $\epsilon \ll 1$, and the energy width $\delta =0$]. It is classified as diabatic if instead of the ground subspace a low-energy subspace $\mc{C}$ is preserved for all $t\in[0,T]$, in the sense of Theorem~\ref{th:JRS}, with non-zero energy width $\delta =\mc{O}(1)$.

\subsection{Coherent and $\mc{C}$-coherent \textit{vs} weakly and strongly decoherent}
\label{sec:C-coherent}
We also classify an implementation of TF-HI according to whether it is coherent or decoherent. The term ``decoherent" could refer to a wide range of decoherence effects which may be present, giving rise to different computational models.  
A common distinction is then whether the system decoheres in the computational basis (for which the state of the system is a mixture over classical states and genuine quantum effects are unlikely or impossible) or in its energy eigenbasis (which may still yield a quantum algorithm since energy eigenstates can be highly entangled in the computational basis).\footnote{The subject of adiabaticity in open quantum systems and its relation to AQC is rich and complicated, and we make no attempt to cover it here in any detail. See Refs.~\cite{childs_robustness_2001,PhysRevLett.95.250503,amin_decoherence_2009,oreshkov_adiabatic_2010,Qiang:13,Avron:2012tv,Albash:2015nx,Venuti:2015kq,Venuti:2018aa} for some entries into this literature.}

In the TF-HI setting there are therefore several relevant notions of coherence. The first is the usual notion of fully coherent, unitary evolution of a closed quantum system. In this case there is coherence also between energy eigenstates. 
The second notion is that of each energy eigenstate being a coherent superpositions of computational basis states, without requiring coherence between energy eigenstates, which is obviously weaker than the first notion. This latter notion of coherence arises in the weak coupling limit (WCL) to the environment, so we call it weakly-decoherent. 
More specifically, we say the TF-HI dynamics are weakly-decoherent if the energy eigenstates remain coherent superpositions of computational basis states over a timescale longer than $T$;\footnote{Note that the commonly used $T_2$-time is a measure of the dephasing time of a single qubit, and is quite different from the time-scale being discussed here.} otherwise we call the dynamics strongly-decoherent. The latter constitutes the third notion of coherence, and arises in the so-called singular coupling limit. 

In the weakly-decoherent case, coherence between different energy eigenstates is irrelevant. For example, a superposition state between the instantaneous ground and first excited state of $H(t)$ may decay on a timescale that is much shorter than $T$, as long as the ground state itself remains a coherent superposition of computational basis states~\cite{Albash:2015nx}. 

In the strongly-decoherent case, where coherence between computational basis states is lost over a timescale shorter than $T$, there is essentially no hope of performing a meaningful quantum algorithm before all useful quantum effects are washed away, so we do not consider this limit further.

To accommodate DQA, we relax the definition of the coherent case to include a scenario that is intermediate between a perfectly isolated quantum system and the weakly-decoherent case. That is, we will say that:
\begin{definition}[$\mc{C}$-coherence]
\label{def:C-coherence}
The dynamics are $\mc{C}$-coherent if the system is sufficiently weakly coupled to its environment so that it is possible to identify a low-energy subspace $\mc{C}$ inside of which coherent superpositions of energy eigenstates are maintained over a timescale larger than $T$, but superpositions with energy eigenstates outside of $\mc{C}$ may decohere. 
\end{definition}
This is the scenario that will natively enable DQA-based algorithms (to go beyond this we briefly discuss how $\mc{C}$-coherence may be protected using Hamiltonian error suppression methods in Sec.~\ref{sec:HES}). Note that when $\mc{C}$ is just the ground subspace, $\mc{C}$-coherent dynamics reduces by definition to the weakly-decoherent case, and when $\mc{C}$ is the entire Hilbert space we recover the fully coherent case. Note further that $\mc{C}^\perp$ (the subspace of states outside of $\mc{C}$) need not necessarily contain only states higher in energy than those in $\mc{C}$. Namely, in some cases $\mc{C}$ will be defined by a symmetry of $H(t)$, and states in $\mc{C}^\perp$ will be those that do not obey this symmetry. A relevant example we discuss below is qubit-permutation symmetry, where $\mc{C}$ is spanned by all the states that are invariant under qubit permutations.

\subsection{Four forward annealing cases}

In light of the considerations above we discuss four cases, summarized in the first two columns of Fig.~\ref{fig:4-cases}. In all these cases we consider only the standard ``forward" annealing protocol, with $A(t)$ and $B(t)$ monotonically decreasing and increasing, respectively.

\subsubsection{Coherent Forward Adiabatic TF-HI}
The case of unitary, fully coherent adiabatic TF-HI (also sometimes known as quantum adiabatic optimization, or AQO~\cite{Farhi:00,Smelyanskiy:01,Reichardt:2004}) is the most theoretically well-studied TF-HI variant because a sufficient adiabatic timescale is determined by the minimum spectral gap of $H(t)$ above its ground state.  This model is capable of a quadratic speedup in the Grover search problem when using a finely-tuned annealing schedule~\cite{Roland:2002ul,Albash-Lidar:RMP,RPL:10}, but it is not robust to control noise~\cite{Slutskii:2019aa} and many other results are negative.  These negative results include apparently efficient classical simulation by quantum Monte Carlo (QMC)~\cite{sqa1,Brady:2015rc,Mazzola:2017aa}, and also the generic occurrence of the many-body localized phase that these Hamiltonians enter at the end of the anneal, which is associated with an exponential number of avoided level crossings~\cite{Altshuler2010}.   The issue of efficient QMC simulation is not resolved in general; there are known families of problems which can be solved by coherent adiabatic TF-HI in polynomial time, but which create topological obstructions for QMC methods that cause an exponential slowdown~\cite{Hastings:2013kk}.   There has been an effort to realize these obstructions in problem Hamiltonians of interest for optimization and sampling applications~\cite{Andriyash:2017aa,vinci2019path}, as well as quantum simulations of topological phases~\cite{King:2018aa,King:2019aa}, and thereby find a useful performance advantage for coherent adiabatic TF-HI over QMC.   One issue with this approach is that mathematical and complexity theoretic properties of coherent adiabatic TF-HI (i.e., the reliance on ground states of stoquastic Hamiltonians) make the possibility of efficient classical simulation (e.g., by improved future QMC algorithms) more likely than for general quantum dynamics.\\  
Therefore \emph{we do not regard coherent adiabatic TF-HI as among the most promising candidate protocols for achieving a quantum advantage.}  

\subsubsection{Weakly-decoherent Forward Adiabatic TF-HI}
Next, we consider weakly-decoherent TF-HI in the open system adiabatic regime in which all decoherence is caused by coupling to an environment so that the energy levels of the original system remain well-defined for the duration of the anneal time $T$.  The state of the system then remains close to the instantaneous thermal state of $H(t)$~\cite{Venuti:2015kq}.  This causes the algorithm to suffer from a similar complexity theoretic limitation as coherent adiabatic TF-HI, which is that the instantaneous thermal state of a stoquastic Hamiltonian is also amenable to QMC methods.  Again it remains unresolved whether QMC can efficiently simulate these states in all relevant cases, but several rigorous thermal state simulations have been developed in recent years~\cite{Harrow:2019aa,kuwahara2019clustering,crosson2020classical}, and QMC has been used to simulate transverse-field spin glasses with thousands of qubits~\cite{Albash:2017aa,Mandra:2017ab,DW2000Q}.\\ 
Therefore \emph{we do not regard (weakly-)decoherent adiabatic TF-HI as a promising candidate for a quantum advantage.}

\subsubsection{Weakly-decoherent Forward Diabatic TF-HI}
If the system decoheres in the energy eigenbasis this may be beneficial since excitations caused by diabatic transitions can in principle be later relaxed by open system thermal processes (in addition to diabatic transitions that can also de-excite the system)~\cite{TAQC,DWave-16q}.   Since simulating the full open system dynamics is classically intractable for systems larger than about $15$ qubits, our limited understanding of this computational model so far comes from the results obtained on D-Wave devices, which correlate reasonably well with path-integral QMC and spin vector Monte Carlo (SVMC), a fully classical model of interacting compass needles~\cite{SSSV}. 

The D-Wave devices have been used to thoroughly explore the weakly-decoherent diabatic version of TF-HI (arguably, the weak-coupling limit assumption is in fact overly generous), and ultimately the results of that exploration appear at this point to be negative: while a scaling advantage over limited classical algorithms like simulated annealing has been found~\cite{Albash:2017aa}, no scaling advantage of the D-Wave devices over state-of-the-art classical competition has so far been shown for any optimization problem or related application of interest~\cite{speedup,Mandra:2017ab}.  In addition to a lack of speedup, a surprising empirical finding is the continued success of classical algorithms like QMC (and to a lesser degree SVMC~\cite{Albash:2014if,q-sig2,Mishra:2018}) in simulating the outputs of the D-Wave device, in spite of the general quantum dynamics that would be expected in the limit of a very short anneal time.  
Therefore, given the currently available evidence, we cannot regard weakly-decoherent diabatic TF-HI as a promising candidate for a quantum advantage.

However, we hasten to add that this conclusion may be more a reflection of the D-Wave devices so far operating in the weakly-decoherent regime \emph{with a significant level of control noise}, than the true computational power of the weakly-decoherent regime. Indeed, such control noise is known to be highly detrimental to the success of TF-HI as an optimization algorithm~\cite{Zhu:2015pd,Albash:2019ab}, and significant performance improvements have been demonstrated with the simplest error suppression strategy compatible with the D-Wave devices~\cite{Pearson:2019aa}.\\
Thus \emph{we leave open the possibility that weakly-decoherent diabatic TF-HI can be a promising candidate for a quantum advantage.}

\subsubsection{Coherent and $\mc{C}$-coherent Forward Diabatic TF-HI}
\label{CHTF-HI}

Finally we turn to coherent versions of diabatic TH-HI, which we regard as the most promising versions of the algorithm. We consider both the fully coherent and $\mc{C}$-coherent versions. The fully coherent version---which assumes a perfectly isolated quantum system---is of course an idealization, but is crucial to consider nonetheless as a theoretical construct, in the same vein as quantum algorithms such as Shor's algorithm~\cite{Shor:94} in the closed system version of the gate-model. The $\mc{C}$-coherent version straddles the purely theoretical and more practical realms, though it is admittedly still a limit that appears to be difficult to reach in practice, and very little is known about experimental conditions under which it would arise. One of our motivations here is to present this as a challenge to the community, and stimulate discussion regarding error suppression and correction methods that would allow $\mc{C}$-coherent dynamics to be realized, even approximately.

While the three cases discussed above all appear to be classically simulable by QMC and SVMC (in practice, even if the theory is unresolved) there are no candidate algorithms for efficiently simulating coherent evolutions with diabatic transitions in a transverse-field system, due to the nonequilibrium nature of such systems. Even if we turn to brute-force exponential time simulation methods, preliminary indications suggest that classical simulation of this time evolution can be even more demanding than simulating random quantum circuits, due to the need to track avoided crossings at very precise locations~\cite{Mandra-talk}.   

It is worth emphasizing that classical simulation by QMC involves two essential ingredients: (1) the Hamiltonian must be stoquastic, and (2) the system must be in an equilibrium state (ground state or thermal).  With full controllability, non-equilibrium dynamics of transverse-field Ising models can generate universal sets of gates (indeed some leading platforms such as trapped ions and Rydberg atoms use gates generated by effective Ising models~\cite{Bernien:2017aa,Zhang:2017aa}).  Even with more limited controls, AQC in excited states of stoquastic Hamiltonians can be universal~\cite{Jordan:2010fk}, and the universality of quantum walks shows that even a time-independent stoquastic Hamiltonian (corresponding to the adjacency matrix of a graph) can generate universal quantum dynamics~\cite{PhysRevLett.102.180501}. However, all of these universality protocols come with substantial overheads that make them relatively impractical. Therefore we do not propose spending the effort to implement these protocols, but rather mention them here to justify our confidence in the classical intractability of simulating coherent diabatic TF-HI.

\begin{figure*}[t]
\subfigure{\includegraphics[width=1.2\columnwidth]{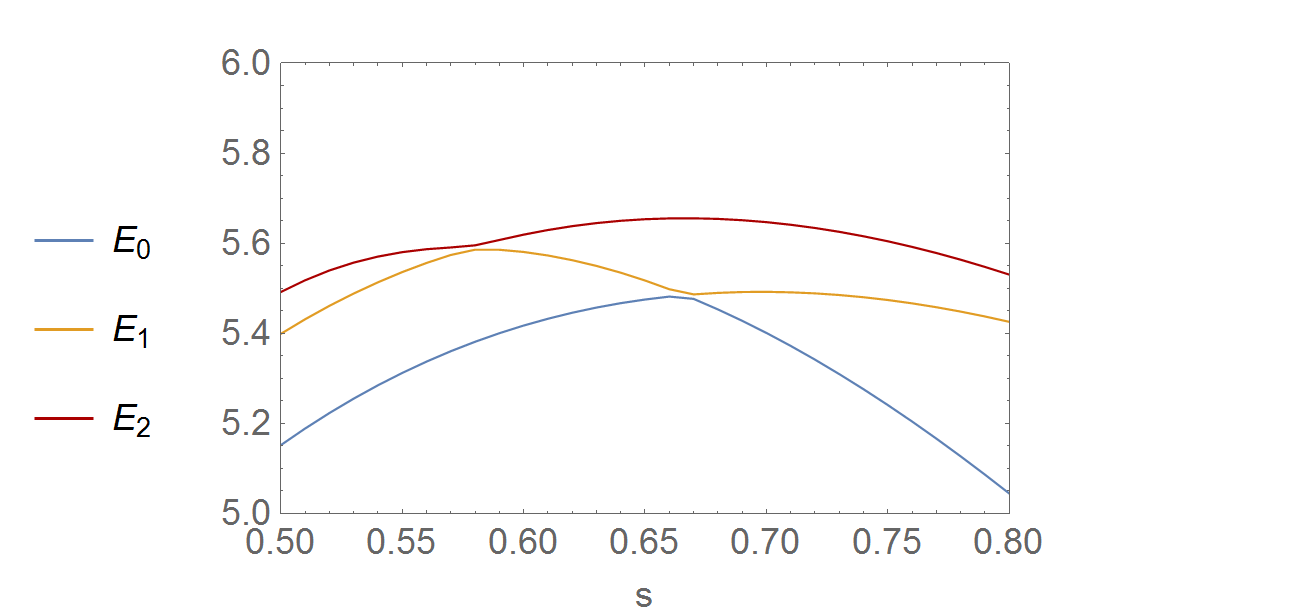}\label{fig:AC}}
\subfigure{\includegraphics[width=.8\columnwidth]{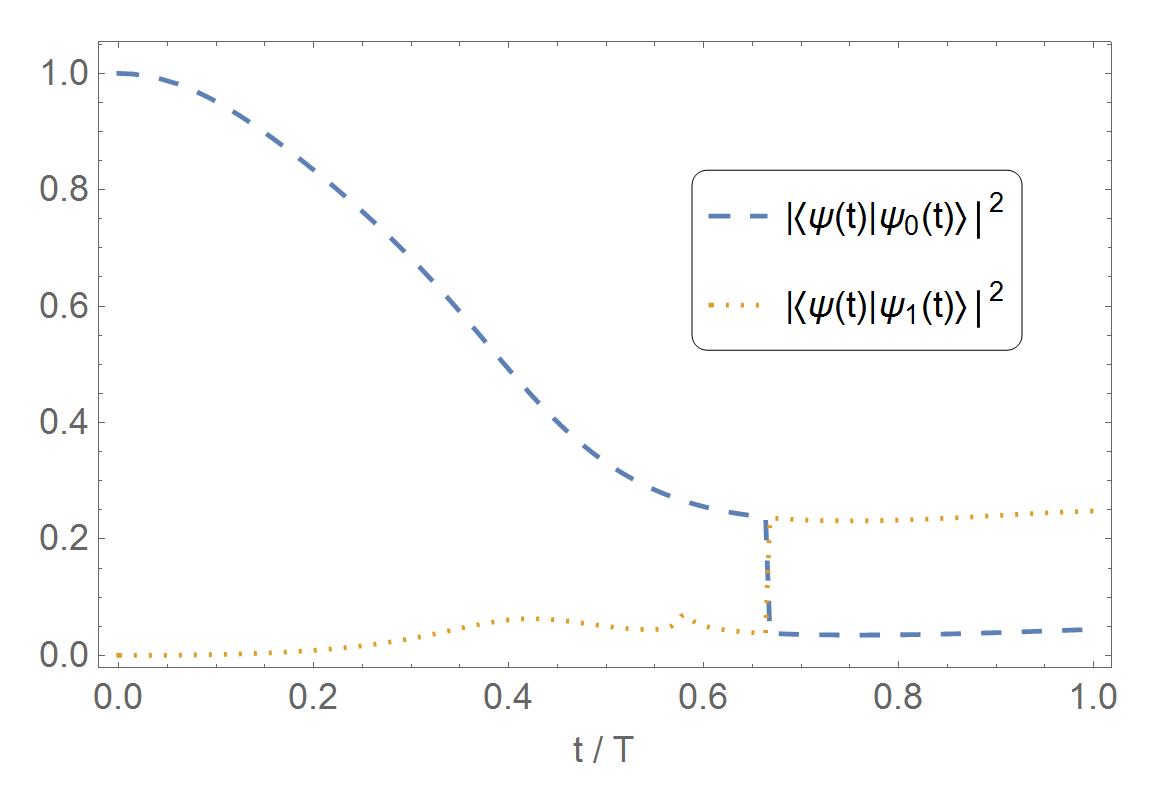}\label{fig:OE}}
\caption{A numerical study of MAX-2-SAT at $n = 20$ qubits.
Left: The lowest 3 energy levels of a particular hard instance in the study~\cite{crosson2014different}, focused on values of $s=t/T$ near two avoided crossings and small gaps between these eigenstates.
Right: The squared overlap of the instantaneous state $|\psi(t)\rangle$ with the ground state $|\psi_0(t)\rangle$ and first excited state $|\psi_1(t)\rangle$, for a choice of the total evolution time $T$ that corresponds to DQA.  The populations in these two eigenstates are exchanged at the avoided crossing between these energy levels.  The key to the diabatic advantage (compared with adiabatic protocols) is the presence of amplitude in the first excited state prior to the avoided crossing.  Note the small bump in amplitude in the first excited state around $s = 0.58$, which corresponds to the avoided crossing between the 1st and 2nd excited energy levels shown in the left panel. Source: Ref.~\cite{crosson2014different}.}
\label{fig:DQA}
\end{figure*}

Besides avoiding the threat of classical simulation, coherent diabatic TF-HI also offers the most intriguing possibilities for algorithmic exploration of prospective enhancement.  A numerical study of the MAX-2-SAT problem at $n = 20$ qubits found that for all of the hardest random instances, the optimal annealing times $T$ were orders of magnitude less than the adiabatic timescale (meaning that diabatic transitions sped up the performance dramatically)~\cite{crosson2014different}.   An understanding of this phenomenon was developed based on the tendency to excite the system early on in the anneal, so that some fraction of the probability returns to the ground state after diabatic transitions de-excite the system. Figure~\ref{fig:DQA} illustrates the low-energy spectrum for one of these hard instances, as well as the overlap of the state of the system with the ground state and first excited state throughout a DQA protocol.    The avoided crossing that creates a small spectral gap between the ground and first excited state near $s = 0.66$  causes the population in the ground state and first excited state to be exchanged at that point, unless the total evolution time $T$ is increased by several orders of magnitude (the time scale in Fig.~\ref{fig:DQA} is $T = 10$, whereas the time scale needed to pass through the minimum gap adiabatically is approximately $T = 10^4$).  Proceeding diabatically early in the anneal excites $\sim 5\%$ of the population into the first excited prior to the avoided crossing, which then becomes a $\sim 5\%$ probability of finding the ground state at the end of the anneal.  Contrast this with a system that evolves adiabatically prior to the avoided crossing; in this case there will be a negligible overlap with the first excited state prior to the crossing, and a negligible overlap with the ground state afterwards.  In these hard instances at $n = 20$ bits, the probability of finding the ground state when $T = 10$ is always $10^2 - 10^3$ times larger than the probability of finding the ground state when $T = 100$ (a time scale which is large enough to be adiabatic at all points other than the avoided crossing).  This study was limited to $n\leq 20$ because exponential-time simulation of the Schr\"odinger dynamics was required, and so the scaling of this effect with system size remains unexplored. At larger system sizes one would expect a more complicated energy spectrum in the late part of the anneal, with a more complicated competition between diabatic excitation and de-excitation, but this study suggests that shorter anneal times with coherent diabatic TF-HI should be investigated at intermediate scales.  Indeed, a closely related study of similar instances up to $n=28$ was reported in Ref.~\cite{Wecker:2016lh}, though the primary context was QAOA. Diabatic transitions were identified as the mechanism explaining the success of machine-learned schedules that improved performance relative to linear schedules for the QAOA angles.

There is also evidence that initial state preparation can improve the performance of annealing: the same study which found all of the hardest instances of Max 2-SAT at $n = 20$ qubits to be improved by shorter anneal times also found that a similarly dramatic improvement in the residual energy (solution error) at a fixed anneal time could be obtained by initializing the system in a  first excited state of the transverse field~\cite{crosson2014different}.\footnote{It is not clear whether producing excited states of product form, e.g., flipping a single spin to minus, $|++...+-+...++\rangle$, would provide anything worthwhile.  What~\cite{crosson2014different}  did was degenerate first-order perturbation theory on the first excited state space of the transverse field, perturbed by the problem Hamiltonian (which can be done efficiently in general), and then initialize the system in an entangled superposition of $+/-$ states that corresponded to the true first excited state.}  
The explanation is familiar: these instances all had avoided crossings late in the annealing schedule that created the opportunity to de-excite the system by a diabatic transition.  This mechanism explains why creating the initial excitation directly worked just as well as using a short anneal time to create the initial excitation. As long as a transverse-field Ising system is initialized in the excited state of the transverse field, and if it remains sufficiently coherent to avoid environmental thermalization, then the system will in general be out of equilibrium and classically intractable to simulate.     
  \textit{While excited states and shorter anneal times are both ways to take advantage of diabatic de-excitation later in the anneal, they in general produce distinct evolutions, and so adding the capability for initializing excited states increases the algorithmic expressiveness at a relatively low cost of control complexity.}

A final example to motivate algorithmic exploration with $\mc{C}$-coherent diabatic TF-HI comes from the study of permutation-symmetric toy problems, whose Hamiltonian has the general form of Eq.~\eqref{eq:h}, with $H_Z=\sum_x f(|x|)|x\rangle\langle x|$, where $x\in\{0,1\}^n$ (a bit-string) and $f$ a cost function that depends only on the Hamming weight $|x|$ (the number of $1$'s in $x$).  Here a simple problem called the ``Hamming weight with a spike" was proposed which can be solved classically in $O(1)$ time using global methods, but which uses a spike barrier [$f(|x|)=|x|$ everywhere except $f(n/4)=n$] to create a local minimum that deceives certain local search algorithms like classical simulated annealing.  By analyzing the minimum spectral gap for this problem it was shown that coherent adiabatic TF-HI can tunnel through this local minimum in polynomial time, though it would take time at least $O(n)$ to do this adiabatically~\cite{Farhi-spike-problem}. However, it was shown that $\mc{C}$-coherent diabatic TF-HI finds the global minimum of this system in $O(1)$ time~\cite{Muthukrishnan:2015ff}. Here $\mc{C}$ is the symmetric subspace, i.e., the subspace of states invariant under qubit permutations.  This is an extreme example in which the anneal time remains constant as the system size grows, described by a ``diabatic cascade" up into and then later down out of the excited energy spectrum.   While the symmetry in the problem contributes substantially to this effect~\cite{Brady:2017aa}, it also demonstrates a key point about the proposal of algorithmic exploration.   The symmetric subspace of $n$ qubits has Hilbert space dimension $n + 1$ which makes classical simulation possible with thousands of qubits. Therefore permutation-symmetric problems are the only place where it has been possible to numerically study ($\mc{C}$-)coherent diabatic TF-HI, and in this tiny corner of parameter space where exploration was possible a dramatic speedup effect was found.  Another pleasing aspect of permutation-symmetric problems is that the symmetric subspace becomes a decoherence-free subspaces when the coupling to the environment is through collective degrees of freedom~\cite{Zanardi:97c,Lidar:1998fk}. When this is not the case, dynamical decoupling can be used to generate the required collective decoherence conditions~\cite{Viola:2000:3520,PhysRevLett.100.160506}. This provides a mechanism to enforce $\mc{C}$-coherent dynamics through error suppression instead of invoking the full arsenal of fault tolerance.\\
Thus, \emph{we regard coherent and $\mc{C}$-coherent diabatic TF-HI as promising candidates for a quantum advantage.}

\section{Protocols with additional control requirements}
\label{sec:alg2}

The difficulty of achieving a quantum speedup with coherent adiabatic TF-HI was noted by numerous authors (e.g., Refs.~\cite{vanDam:01,Altshuler2010,Reichardt:2004,Jorg:2010qa,Laumann:2012hs,Laumann:2015sw,Knysh:2016iq}), and a variety of modifications were accordingly proposed based on introducing additional control requirements. Some such modifications were reviewed in detail in Ref.~\cite[Sec.~VII]{Albash-Lidar:RMP}, and in this section we primarily (but not exclusively) consider more recent developments. 

Quite generally, modifications to TF-HI are algorithms based on more general state preparation capabilities and more general time-dependent control than implied by the Hamiltonian in Eq.~\eqref{eq:h}. The more recent protocols include reverse annealing, rapid quenches, and oscillations in the strength of $A(t), B(t)$ which resemble to some degree the gate sequences used in QAOA.  A general feature of post-TF-HI algorithms is that they introduce additional parameters which can be tuned to search for a speedup, in contrast with TF-HI where the anneal time $T$ is the only free parameter (though environmental engineering can already introduce additional parameters to TF-HI~\cite{theis2018gapindependent}).  The advantage of having more parameters or settings with which to run the algorithm is that it casts a wider net in the search for speedup, while the disadvantage is the need to guide choices of these settings,  a task which can in some extreme cases take on the NP-hard complexity from the original problem.   Therefore we seek to maximize algorithmic expressiveness while minimizing the complexity required for control and parameter selection. 
Diabaticity is an integral aspects of most of the protocols we describe below, but we shall also see examples with purely adiabatic protocols.

\subsection{Reverse Annealing}

The term ``reverse annealing"~\cite{Den2017,DW2017} is somewhat of a misnomer: it does not mean that the anneal proceeds from $H_Z$ to $H_X$, but rather that the latter evolution is concatenated with a standard forward anneal. 
Thus, the evolution starts in the ground state of a Hamiltonian $H_{Z,1}$ that is diagonal in the computational basis, interpolates to an off-diagonal Hamiltonian (the ``reverse" part), inverts, and ends with a new diagonal Hamiltonian $H_{Z,2}$ (the ``forward" part). In the coherent adiabatic version of this protocol the system remains in the ground state at all times, while in the diabatic version one attempts to exploit excitations during the evolution. An early example of the use of reverse annealing was the tunneling spectroscopy experiment demonstrating entanglement in the D-Wave processors~\cite{DWave-entanglement,Albash:2015pd}.

\subsubsection{Diabatic reverse annealing}

\paragraph{Provable exponential speedup in the $\mc{C}$-Coherent case.}
There is a strong theoretical result supporting enhancement via $\mc{C}$-coherent diabatic reverse annealing with a stoquastic Hamiltonian~\cite{Somma:2012kx}. This model natively (i.e., without the overheads that accompany more generic circuit-to-Hamiltonian constructions) reproduces an exponential speedup that was first discovered in the quantum walk model for a problem called ``glued-trees"~\cite{childs2003exponential}. The Hamiltonian is of the form 
\beq
H(s) = (1-s)H_0 + s(1-s)\mc{A} + sH_1 ,
\label{eq:GT}
\eeq 
where $H_0$ and $H_1$ are both diagonal in the computational basis and $\mc{A}$, the adjacency matrix of the glued-trees graph, is off-diagonal and replaces the usual transverse field.  The problem is itself contrived, control-noise sensitive~\cite{Muthukrishnan:2019aa}, and requires oracle access to the description of the graph via $\mc{A}$, but the reason we consider it to be a promising example is that (1) the exponential speedup is achieved by a stoquastic Hamiltonian that is interpolated in such a way as to be very slightly out of equilibrium, and (2) the diabatic transitions between the ground state and first excited state are an essential part of this speedup. This result requires coherence between these two energy eigenstates but not more, so it is an example of the $\mc{C}$-coherent case, with $\mc{C}$ being the subspace spanned by the two lowest energy eigenstates. 

\paragraph{Generalization to the weakly-decoherent case.}
Moreover, if we can prepare the quantum thermal state $\rho_\beta(s)$ for polynomially small temperature $1/\beta$ at each value of $s$ [Eq.~\eqref{eq:beta} below], then sampling this Gibbs state at the final $s$ would suffice to solve the problem as well, i.e., the algorithm also works in the weakly-decoherent setting.  To justify this last claim in detail, we refer to Ref.~\cite[Fig.~2]{Somma:2012kx}, which explains that the dimensionless time $s$ can be divided into 5 regions by defining points $s_1 <  s_2 <  s_3 < s_4$ at which the eigenvalue behavior changes.  The two lowest energy levels $E_0(s), E_1(s)$  satisfy
\begin{align}
E_1(s) - E_0(s) &\geq c / n^3 \ , \quad s \in [0,s_1]\cup [s_2,s_3]\cup [s_4,1] \notag \\
E_1(s) - E_0(s) &= \mathcal{O}(2^{-n/2}) \ , \quad s \in [s_1,s_2] \cup [s_3,s_4]
\end{align}
for some constant $c>0$.  We can fix the inverse temperature to be polynomially small in such a way that the thermal state $\rho_\beta(s)$ is arbitrarily close to the ground state at the end of the anneal.  Let $H$ be a Hamiltonian on $n$ qubits with ground state $|E_0\rangle$ of energy $E_0$ and a gap $\Delta=E_1-E_0$ to the first excited state, then setting $\beta = \Delta^{-1}\left( n \ln(2) + \ln(\delta^{-1}) \right)$ suffices for the partition function $Z = \Tr e^{-\beta H}$ to satisfy 
\beq
1 \leq Z =  \sum_{i = 0}^{2^n -1} e^{-\beta E_i} \leq 1 + e^{-\beta \Delta} 2^{n-1} = 1 + \frac{\delta}{2} ,
\eeq
so that:
\bes
\begin{align}
&\|\frac{1}{Z}e^{-\beta H} - |E_0\rangle \langle E_0| \|_1 \leq \frac{1}{Z}\left \|\sum_{i =1}^{2^n -1} e^{-\beta E_i} |E_i\rangle \langle E_i| \right \|_1 \notag \\
&\quad + \| (1-Z)|E_0\rangle \langle E_0| \|_1 \leq \frac{1}{Z}\frac{\delta}{2} +(Z-1) \leq {\delta} ,
\end{align}
\ees
where $\|A\|_1 \equiv \Tr\sqrt{A^\dag A}$ (the trace norm).
Therefore the density matrix $\rho_\beta = e^{-\beta H}/Z$ is within trace-norm distance $\delta$ of the ground state $|E_0\rangle \langle E_0|$.  It follows that taking the inverse temperature 
\beq
\beta = (n^3/c) \left( n \ln(2) + \ln(\delta^{-1}) \right)
\label{eq:beta}
\eeq 
suffices to make the thermal state of the glued-trees Hamiltonian arbitrarily close to the ground state whenever $s\notin [s_1,s_2]\cup [s_3,s_4]$.  When the gap between the ground state and first excited state becomes exponentially small, the thermal state will contain the nearly uniform mixture of the ground and first excited states.  The point of all this is that if we could guarantee the efficient and accurate preparation of $\rho_\beta(s)$ for all $s \in [0,1]$, and Eq.~\eqref{eq:beta} holds, then this result would provide an oracle separation between classical computing and the model which follows the instantaneous thermal state of a stoquastic Hamiltonian.  This weakly-decoherent version of the glued trees problem is an important venue for a future investigation; we expect that the possibility of the required preparation of $\rho_\beta(s)$ can be proven within the setting of the open-system adiabatic theorem~\cite{Venuti:2015kq,Venuti:2018aa}.\\
With the strong caveat that a physical implementation of the glued trees problem requires many-body interactions, \emph{we regard $\mc{C}$-coherent and even weakly-decoherent diabatic reverse annealing as promising candidates for a quantum advantage.}

\subsubsection{Coherent adiabatic reverse annealing}

\paragraph{Provable superpolynomial speedup.}
It is within the reverse annealing model that the first provable superpolynomial speedup using \emph{stoquastic adiabatic} computation was very recently realized, in the Hamiltonian oracle model~\cite{hastings2020power}. In close analogy to Eq.~\eqref{eq:GT} of the glued-trees algorithm~\cite{Somma:2012kx}, the Hamiltonian path $H(s)$ corresponds at each $s$ to the weighted adjacency matrix $\mc{A}(s)$ of a graph, and the graph is specified by an oracle in the sense that for any vertex, a query to the oracle returns the matrix elements of $\mc{A}(s)$ corresponding to neighbors of that vertex.  In this way the oracle only reveals the local structure of the graph.  The computational problem is to determine a global property of the graph---whether or not a large cycle is present---using only the queries to the oracle that reveal the local structure.  The proof of a super-polynomial speedup for this problem involves showing that stoquastic adiabatic computation solves it in polynomial time, and proving that no classical algorithm can solve the problem using fewer than $n^{\Omega(\log(n))}$ queries to the oracle. 

However, while the algorithm is suitable for gate-model Hamiltonian simulation of the adiabatic algorithm, it is unsuitable for analog implementation with a \emph{local} Hamiltonian, since it requires many-body interactions. In this regard it is similar to the glued trees algorithm and the adiabatic Grover's algorithm~\cite{Roland:2002ul}. Still, the result is a significant advance since it combines adiabaticity (unlike glued trees) and a superpolynomial speedup (unlike Grover) with stoquasticity, and constitutes the strongest evidence to date in favor of the prospect of a quantum speedup in the coherent adiabatic model. \\
Thus, \emph{we regard coherent adiabatic reverse annealing as a promising candidate for a quantum advantage} (caveated similarly to the glued trees algorithm).

\begin{figure*}[t]
\subfigure{\includegraphics[width=.29\textwidth]{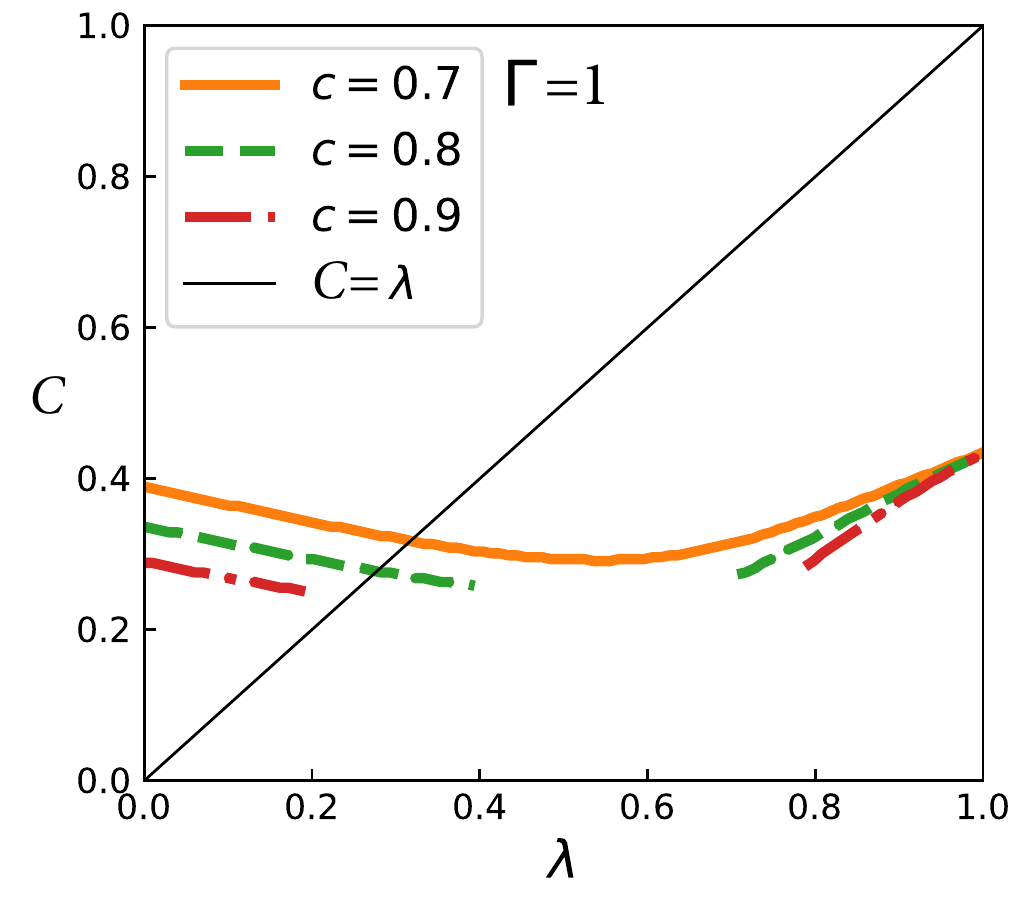}\label{fig:p3G1}}
\subfigure{\includegraphics[width=.29\textwidth]{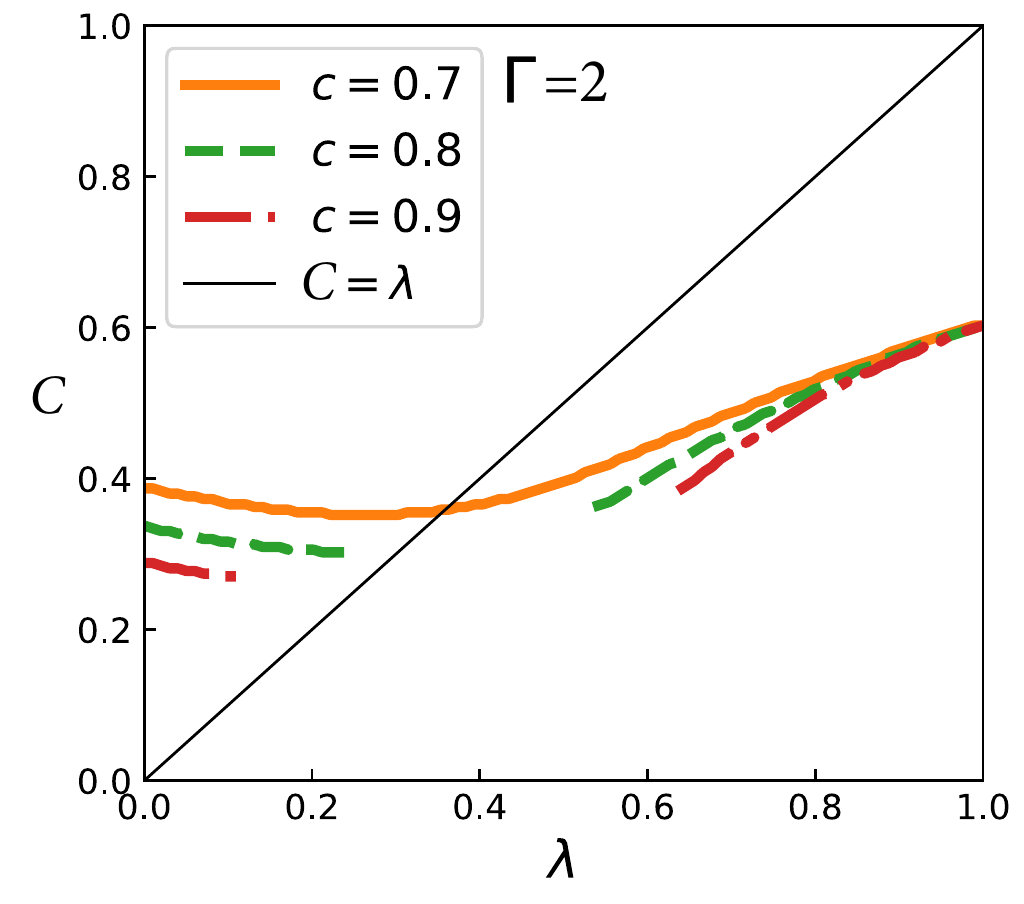}\label{fig:p3G2}}
\subfigure{\includegraphics[width=.34\textwidth]{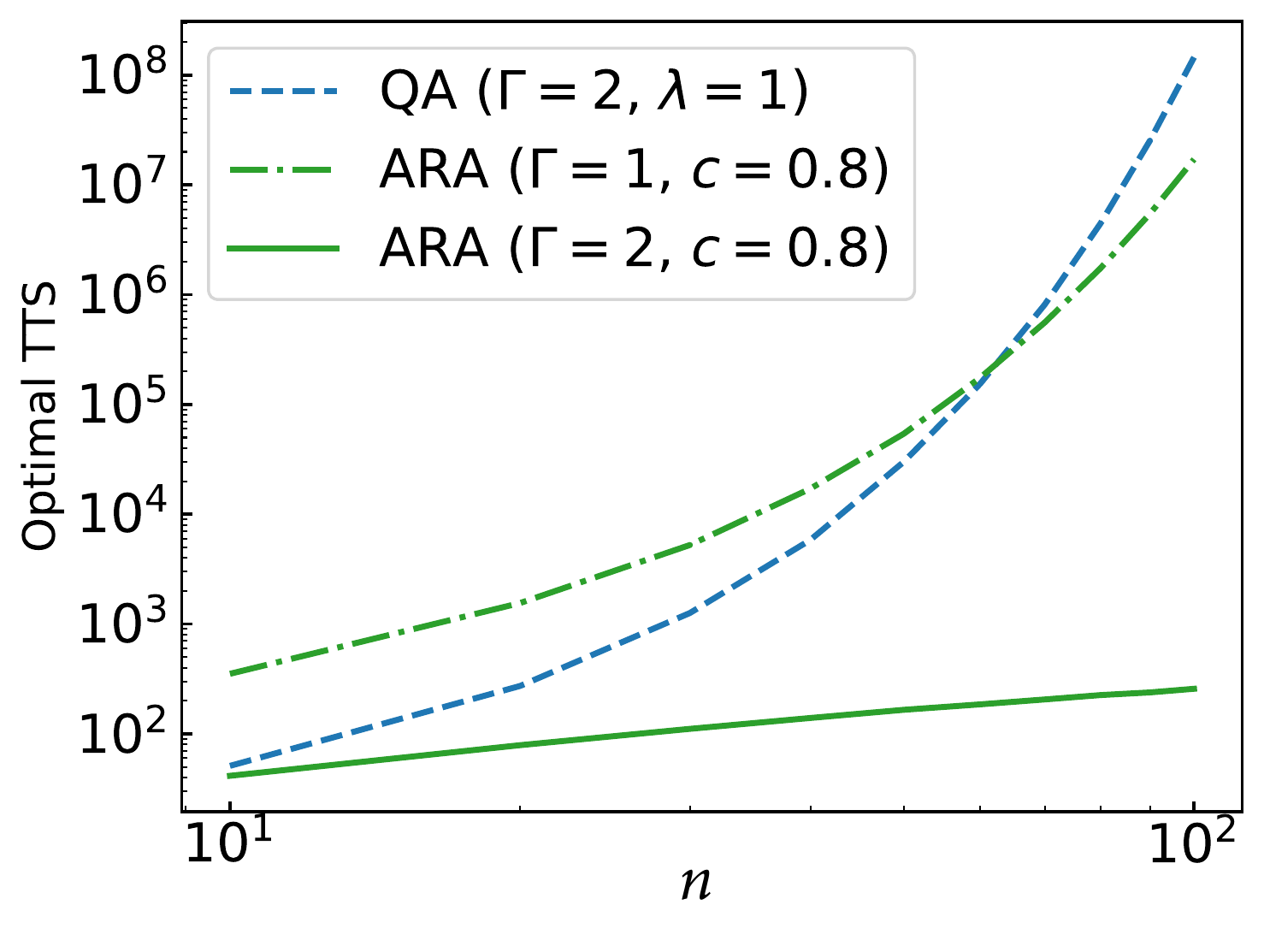}\label{fig:TTS}}
\caption{
Left and middle: Phase diagrams in the \(C\)-\(\lambda\) plane for \(p=3\) for different values of the amplitude of transverse field: $\Gamma=1$ (left) and $\Gamma=2$ (right). Curves indicate first order quantum phase transitions. The parameter $c=\frac{1}{n}\sum_i \epsilon_i$ is the magnetization of the initial state; the ground state has $c=1$. A path that avoids the first order transition is possible for sufficiently large $c$ (depending on the transverse field magnitude $\Gamma$), i.e., when the initial state has significant overlap with the ground state. 
Right: Size dependence of the optimal time-to-solution (TTS) of forward TF-HI (denoted ``QA", blue dashed line; the vertical $\lambda\equiv 1$ path), \textit{vs} the diagonal $C\equiv \lambda$ path of ARA with $\Gamma=1$, $c=0.8$ (green dash-dotted line) and ARA with $\Gamma=2$, $c=0.8$ (green solid line). As seen in the left and middle phase diagrams, ARA with $\Gamma=1$, $c=0.8$ encounters a first order transition whereas $\Gamma=2$, $c=0.8$ does not. This difference in statics is reflected in the dynamics as the exponential and polynomial dependence of the TTS. Source: Ref.~\cite{Yamashiro:2019aa}.}
\label{fig:ARA}
\end{figure*}

\paragraph{Softening a quantum phase transition.}
First order quantum phase transitions are canonical examples of failures of TF-HI, since they are typically associated with exponentially small gaps, thus incurring exponentially long adiabatic evolutions even for problems that are trivial to solve classically by inspection~\cite{Reichardt:2004,Jorg:2010qa,Laumann:2012hs,Laumann:2015sw}. Adiabatic reverse annealing is one method for circumventing such obstructions to the success of quantum annealing.
Indeed, consider the following generalization of Eq.~\eqref{eq:h}: 
\begin{align}
H(t) &= \Gamma \lambda(t) A(t) H_X + B(t) H_Z + (1- \lambda(t)) C(t) H_{\text{init}} .\notag \\
H_{\text{init}} &= -\sum_{i=1}^n s_i Z_i .
\end{align}
Note that when $\lambda(t)\equiv 1$ we recover $\HTIM(t)$. Here $A(t)$ and $C(t)$ both monotonically decrease to zero while $B(t)$ monotonically increases from zero for $t\in[0,T]$, and $\lambda(t)$ is an additional control parameter satisfying $\lambda(0)=1-\lambda(T)=0$. The additional term $H_{\text{init}}$ is a diagonal Hamiltonian in the computational basis and serves to enforce a given classical ground state $|s_1,\cdots,s_n\rangle$ (with $s_i=\pm1$) as an initial condition. The Hamiltonian path specified by $\lambda(t)$ and $C(t)$ subject to the constraints above thus implements coherent adiabatic reverse annealing (ARA): the system is initialized in the classical ground state of $H_{\text{init}}$, undergoes quantum fluctuations at intermediate times mediated by $H_X$, and (in the adiabatic limit) ends in the ground state of the problem Hamiltonian $H_Z$.

ARA was studied in Ref.~\cite{Ohkuwa:2018aa} using a static, equilibrium statistical mechanics analysis for the $p$-spin model, defined by $H_Z = -\frac{1}{n}(\sum_{i=1}^n Z_i)^p$ for positive integer $p$. While the ground state is trivial to find (the $|0^{\otimes n}\rangle$ state), the model exhibits a first-order quantum phase transition under standard, forward TF-HI for $p\ge 3$~\cite{Jorg:2010qa}. However, by choosing an appropriate path in the $(\lambda,C)$ plane, ARA turns this into a second order transition, where the gap is only polynomially small. The implication that reverse annealing might provide an exponential speedup relative to forward annealing, at least for the $p$-spin model, was confirmed numerically from the dynamical perspective in a study that found that the time-to-solution (TTS) metric scales polynomially in $n$ for ARA, while the lower bound for the TTS of forward QA is exponential in $n$~\cite{Yamashiro:2019aa}; see Fig.~\ref{fig:ARA}. While encouraging, these ARA results rely on some knowledge about the classical solution being built into $H_{\text{init}}$, through the parameter $c=\frac{1}{n}\sum_i s_i$, so it is unclear how well they generalize to hard optimization problems.

\subsubsection{Iterated coherent and weakly-decoherent reverse annealing}
We discussed iterative protocols that attempt to exploit previous knowledge to construct an improved solution in the next annealing run.

\paragraph{Iterated coherent reverse annealing via the Sombrero-AQC protocol.}
Coherent reverse annealing was first proposed as a heuristic protocol (called ``Sombrero-AQC") designed to be used iteratively, as a means to feed a trial solution from one run into the next~\cite{Perdomo-Ortiz:2011fh}. In this protocol the Hamiltonian is diagonal in the computational basis both at $t=0$ and $t=T$ while it is off-diagonal at intermediate times. The standard forward Hamiltonian of Eq.~\eqref{eq:HTIM} is thus modified to
\bes
\label{eq:RA}
\begin{align}
   \label{eq:sombrero}
    H(t) &= A(t)H_X + B(t)H_Z + C(t)H_{\text{init}}\\ 
    H_{\text{init}} &= -\sum_i g_i Z_i ,
\end{align}
\ees
where $A(t)\geq 0$ is sombrero-shaped with $A(0)=A(T)=0$ (symmetric about the inversion point $t=T/2$), $B(t)$ is monotonically increasing from zero, and the initialization schedule $C(t)$ is monotonically decreasing to zero. The local fields $g_i$ determine the classical initial state.

The main idea going beyond forward TF-HI is to introduce an iteration. Namely, if the state $|\psi(T)\rangle = |s_1,\cdots,s_n\rangle$ (with $s_i=\pm1$) is not the ground state of $H_Z$ (due to a diabatic excitation) then this state can become the ground state of a new diagonal Hamiltonian $H_{\text{init}}$, with $g_i = s_i$. The latter is used as the initial Hamiltonian for a new annealing run, resetting the clock to $t=0$ in Eq.~\eqref{eq:sombrero}.

The iteration allows for quantum fluctuations to assist in the search for lower energy classical states by temporarily delocalizing the system while the transverse field $H_X$ is turned on during the reverse evolution stage. Here delocalization is meant in the sense of creating a (nonuniform) superposition over computational basis states, so that the system might tunnel to a new local minimum during the forward evolution stage. 

The general idea of iteration suggested by the Sombrero-AQC protocol is a powerful one, and has given rise to a variety of heuristic hybrid quantum-classical protocols such as quantum parallel tempering, quantum population annealing~\cite{Chancellor:2016ys}, and a quantum-assisted genetic algorithm (where the reverse evolution is viewed as a mutation operator, and recombination and selection are implemented classically)~\cite{king2019quantumassisted}. In all these cases, numerical simulations indicate improved performance relative to the forward TF-HI protocol applied to the same problem instances.

\paragraph{Iterated coherent reverse annealing with a fixed diagonal Hamiltonian.}
One disadvantage of the Sombrero-AQC protocol is that it requires a reprogramming of $H_{\text{init}}$ for every new cycle. 
One can instead set $H_{\text{init}}=0$ and start every reverse annealing cycle from $H_Z$, and a 
random initial classical state (in general an excited state of $H_Z$). However, this protocol was shown to fail to converge to the ground state of the $p$-spin model~\cite{Yamashiro:2019aa}. The reason is that in order for the protocol to work and provide an enhanced probability of finding the ground state after multiple iterations, the probability distribution of the final state after each iteration would have to shift toward lower energy states than the initial state. This condition was found to be violated in the $p$-spin model. 

The weakly-decoherent version of this protocol, however, does work well for the $p$-spin model, as relaxation to the ground state is made possible by included dephasing in the instantaneous energy eigenbasis~\cite{passarelli2019reverse}. The associated thermal relaxation results in a significant increase in the success probabilities, as long as the inversion point value $A(T/2)$ is chosen to be close to or before the avoided crossing value of $A$. This example of thermal relaxation being the mechanism responsible for the success of the protocol raises interesting questions about whether an intermediate regime of quantum-relaxation-assisted, weakly-decoherent iterated reverse annealing can result in a quantum advantage. Results involving mid-anneal pausing, which we describe next, suggest the answer might be affirmative.

\begin{figure}[t]
\includegraphics[width=1.3\columnwidth]{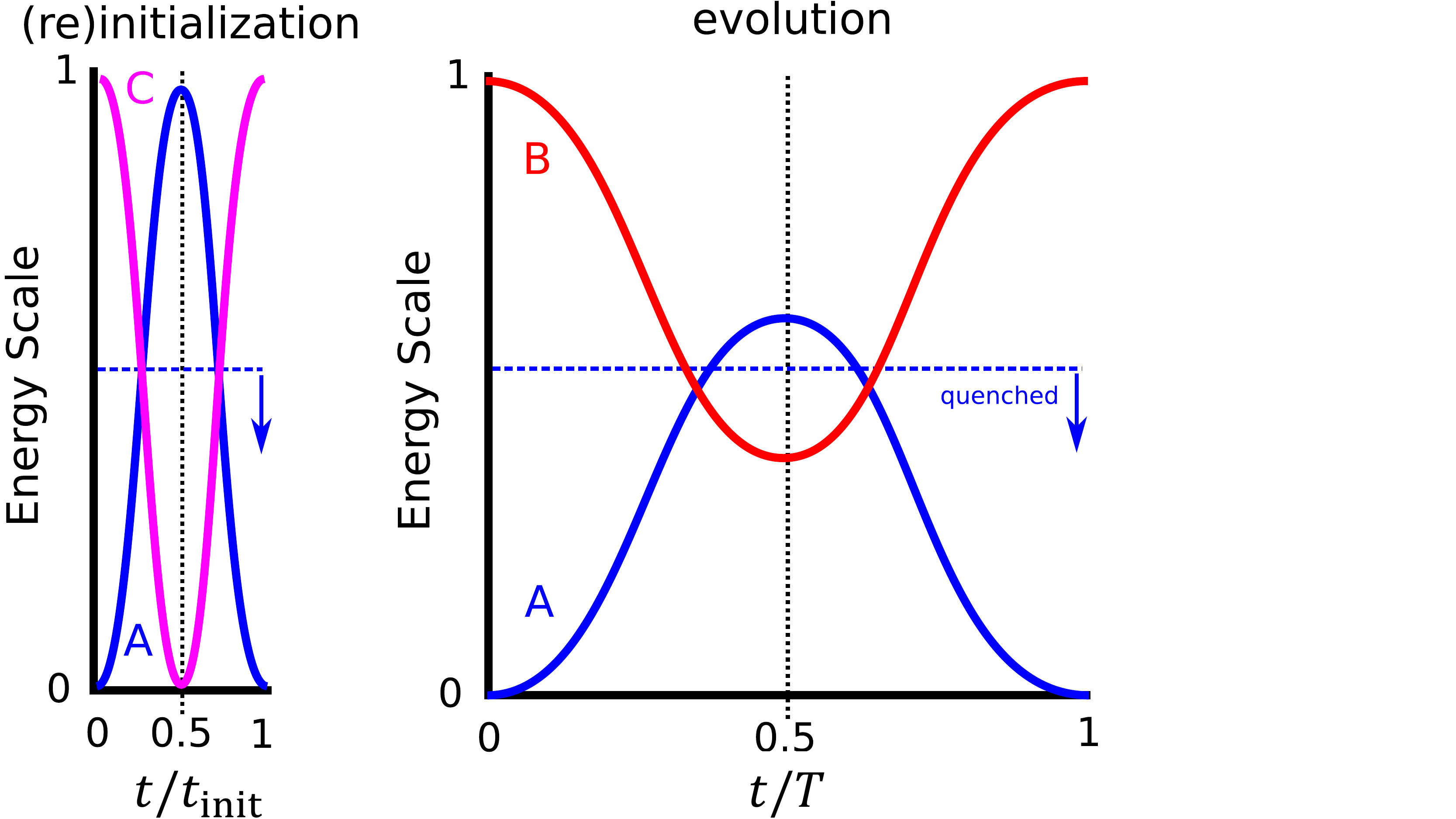}
\caption{The D-Wave reverse annealing protocol. Shown are dimensionless energy scales $A(t),B(t),C(t)\in[0,1]$ in Eq.~\eqref{eq:RA}.  When $A(t)$ is below the dashed horizontal blue line, all dynamics are quenched and the system remains in whatever state it was placed in.  Only when $A(t)$ passes above the horizontal blue line can the system change state. Generally $t_{\text{init}}\ll T$. In the ``reinitialize-state=false" protocol the (re)initialization sequence is applied once, followed by repeated applications of the evolution sequence. In the ``reinitialize-state=true" protocol, the (re)initialization sequence is followed by the evolution sequence, and this is repeated. Pausing corresponds to the insertion of a horizontal segment at $t=T/2$. (Figure courtesy of Richard Harris.)}
\label{fig:IRA}
\end{figure}

\paragraph{D-Wave's iterated reverse annealing protocol.}
An experimental version of iterated weakly-decoherent reverse annealing is possible using the D-Wave 2000Q devices~\cite{DW2KQ}, which also feature a mid-anneal pause~\cite{DW2017}. We sketch the corresponding schedules in Fig.~\ref{fig:IRA}.
The maximum value reached by the transverse field, $A_{\max}=A(T/2)$, plays a crucial role: the dynamics are ``quenched or ``frozen"~\cite{Amin:2015qf} in a classical state for sufficiently small $A(t)$, indicated by the dotted line in Fig.~\ref{fig:IRA}. Thus $A_{\max}$ must be sufficiently large in order for quantum fluctuations to enable an exploration of the system's Hilbert space.

Adding a pause at $t=T/2$ is also possible. Pausing is superficially similar to slowing down near the minimum gap, as in the locally adiabatic Grover schedule~\cite{Roland:2002ul}, but is implemented here in a different context, associated with open system dynamics subject to thermal relaxation. The first study~\cite{marshall_power_2019} 
to empirically test the utility of pausing in optimization demonstrated an improvement in the success probability when a pause was inserted right before the minimum gap point in reverse annealing; this point corresponds to a value of $A_{\max}$ larger than the quenched energy scale, but not so large that memory of the initial state is lost due to the phase transition associated with crossing the minimum gap. Pausing was also used in the entanglement experiment~\cite{DWave-entanglement} and was found to be advantageous in application problems such as portfolio optimization~\cite{venturelli_reverse_2019} and training deep generative machine learning models~\cite{vinci2019path}. That it is beneficial to pause mid-anneal (in the sense of an improved success probability relative to not pausing) was recently rigorously established under certain sufficient conditions on the relaxation rates at the pause point and at the end of the anneal, for a simplified model of a two-level system described in terms of a quantum master equation~\cite{chen2020pausing}. However, it remains to be established that pausing improves optimization performance according to the time-to-solution metric.

Finally, while our focus here is on optimization, it is important to mention that iterated reverse annealing (in the reinitialize-state=false sense of Fig.~\ref{fig:IRA}) was also the protocol used in recent quantum simulations of topological phases using the D-Wave devices~\cite{King:2018aa,King:2019aa}.

\emph{We conclude that all the reverse annealing heuristics mentioned here are potentially promising, and given that they explicitly take advantage of diabatic transitions and very little is known in terms of rigorous results, they are well worth exploring further.}

\begin{figure*}[t]
\subfigure[\ Source:~\cite{pagano2019quantum}.]{\includegraphics[width=1.28\columnwidth]{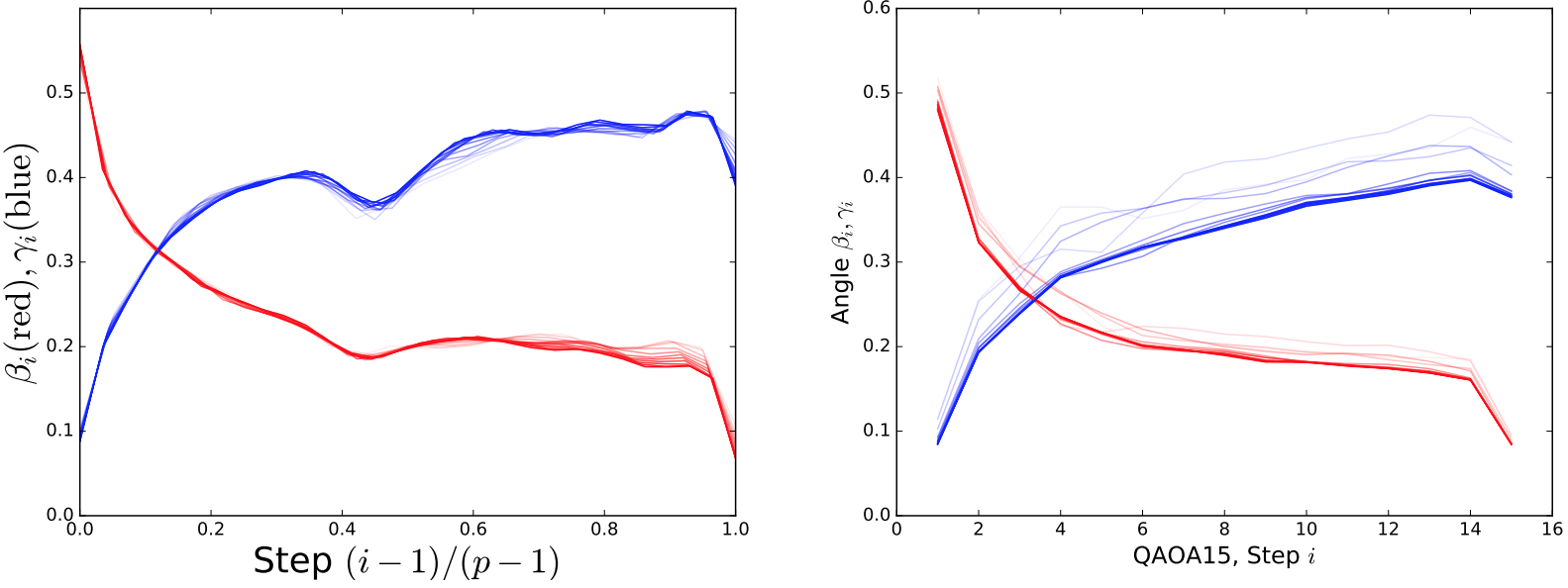}\label{fig:QAOA}}
\subfigure[\ Source~\cite{brady2020optimal}.]{\includegraphics[width=.7\columnwidth]{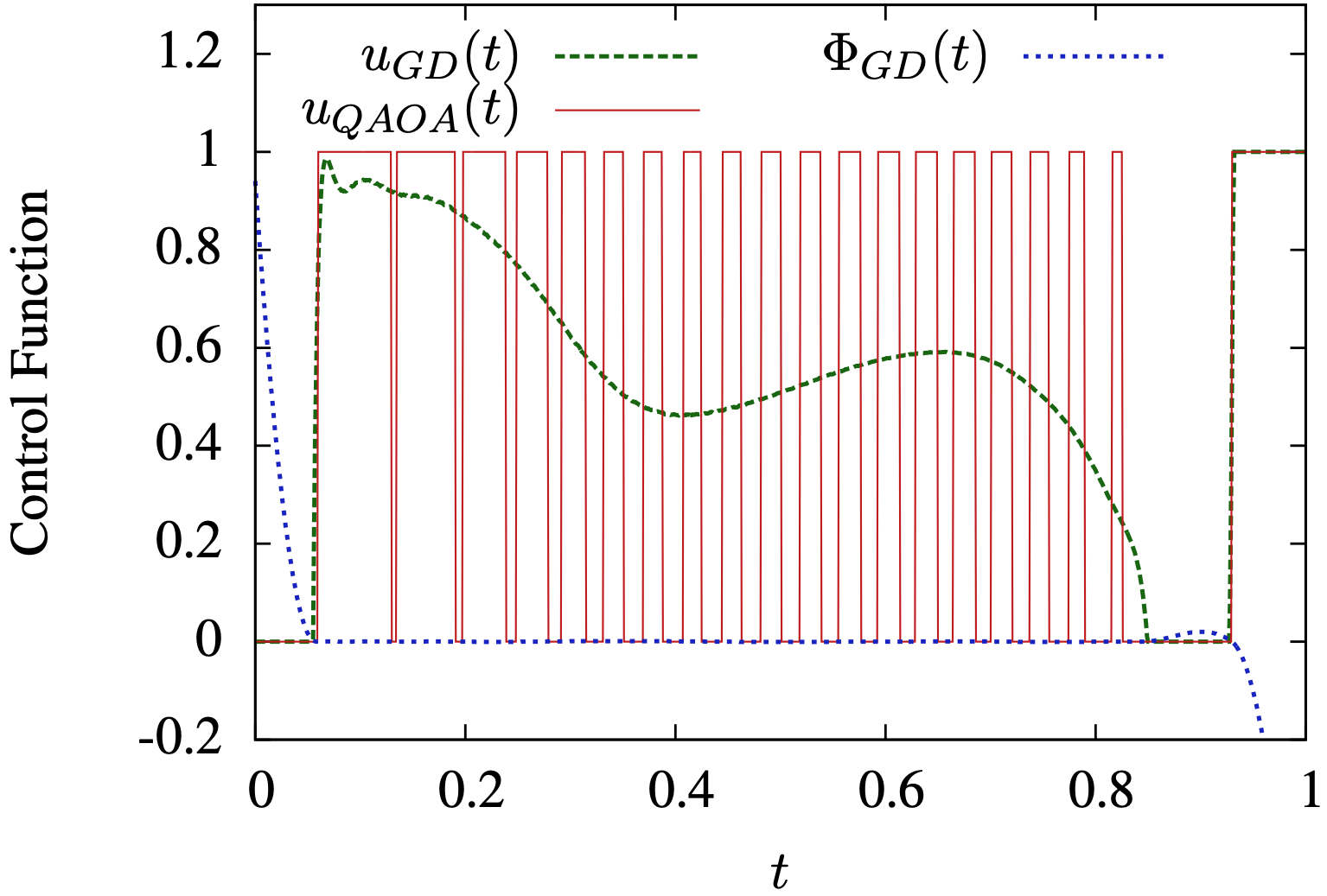}\label{fig:bang-anneal-bang}}
\caption{Optimal QAOA approaches coherent diabatic QA.\\
(a) Convergence in $p$ and $n$. Convergence of optimal angle curves with increasing QAOA layers $p$ (left), and number of qubits $n$ (right). The $p$-convergence plot was generated for $n = 8$ and $p\in[20,30]$, with higher $p$ shaded darker. The $n$-convergence figure was generated for a $15$-layer QAOA and $n\in[4,14]$, with higher $n$ curves shaded darker.\\
(b) Optimal control functions found through either gradient descent ($u_{GD}(t)$) or constrained-time QAOA ($u_{QAOA}(t)$) for a random instance of the MaxCut problem. Also shown is the gradient $\Phi_{GD}(t)$ for the gradient descent method. The $u_{GD}(t)$ schedule outperforms the $u_{QAOA}(t)$ schedule and is described by initial and final bangs, with a smooth schedule in between. Parameters: $n = 8$ qubits, total time $T = 2.0$, $2p = 40$ bangs for
the QAOA method. The schedule is $H(t) = u(t)H_X +(1-u(t))H_Z$.}
\end{figure*}

\subsection{Quantum Walks on a Boolean Hypercube with a Rapid Quench}
The standard framework of continuous-time quantum walks considers the Schr\"odinger equation for a spinless particle hopping on a combinatorial graph, in which case the time-independent Hamiltonian corresponds to the adjacency matrix (or graph Laplacian) of the graph.  In this standard framework the geometry of the graph determines an interference pattern of the walker, which can be exploited to solve unstructured search and is even known to be universal for quantum computation~\cite{PhysRevLett.102.180501}.   

A recent alternative framework fixes the graph to be a Boolean hypercube, with graph Laplacian $L = n I + H_X$ (where as in the case of $\HTIM$, $H_X = -\sum_{i =1}^n X_i$), and weights the vertices according to a classical spin glass cost function $H_C$~\cite{callison2019finding,morley2019quantum,callison2020energetic}. The full time-\emph{independent} Hamiltonian is then taken to be
\beq
H = \gamma L + H_C ,
\label{eq:H_w}
\eeq
where $\gamma$ is the hopping rate, a parameter to be tuned in order to maximize the probability of finding the ground state of $H_C$ after the unitary dynamics $e^{-i T H}$ is applied for sufficiently large $T$.  With the important caveat that it is assumed that the Hamiltonian~\eqref{eq:H_w} is switched on instantaneously at $t=0$ (instantaneous quench, or ``bang"), this model appears well suited for quantum annealing architectures; in fact it requires no time-dependent annealing schedules, only a time-independent Hamiltonian that is kept on for total time $T$. Note that this describes DQA evolution with a Hamiltonian of the transverse-field Ising form. The system is prepared in the usual uniform superposition state in the computational basis (an excited state of $H$; hence there is some similarity to the excited initial state protocol of Ref.~\cite{crosson2014different}), and is measured at $T$ in the same basis. Relaxing the condition that the quench is instantaneous, so that instead the system is prepared in the ground state of $H_X$ and is then rapidly (non-adiabatically) evolved to $H$, would make it practically suitable for quantum annealing hardware implementations.

The study~\cite{callison2019finding} applied this algorithm to $10^4$ instances of the Sherrington-Kirkpatrick (SK) model for $5 \leq n \leq 11$ qubits and also compared the findings against those for the random energy model (in which case $H_C$ is a diagonal matrix with independent and identically distributed Gaussian entries).  The random energy model (REM) represents a problem without any structure, which means that the best solutions are limited to random guessing. While for the REM it was found that the late time success probability as a function of $\gamma$ had a sharp instance-dependent peak, the SK model was relatively robust to heuristic selection of $\gamma$ and found that the late time success probabilities decayed as $P_\infty = 2^{-\alpha n}$ for $\alpha = 0.417  \pm 0.001$.  Despite the heuristic nature of this result (the algorithm has no theoretical guarantee to succeed at late times) and the possibility of finite-sized effects $(n\leq 11)$ this is an encouraging result due to the apparent super-quadratic speedup over brute force search, for which $P_\infty = 2^{- n}$.  

A closely related approach, but using analytical tools, was discussed in~\cite{Hastings2019dualityinquantum} (see also Ref.~\cite{Chancellor2020perspectiveduality}), focusing on the MAX-K-LIN-2 problem with couplings in $\{-1,0,1\}$. It points to the interesting observation that energy conservation is a principle that allows one to obtain lower energy solutions better than a random guess.

We may view these quantum walks results as belonging to the promising category of coherent diabatic forward annealing; see Fig.~\ref{fig:4-cases}. Moreover, the ``bang" assumed at $t=0$ ties these results to recent results on optimal QAOA schedules, which we discuss next.

\subsection{QAOA \textit{vs} coherent diabatic TF-HI}

We briefly compare state-of-the-art results concerning the Quantum Approximate Optimization Algorithm (QAOA)~\cite{farhi2014quantum} and coherent diabatic TF-HI in this subsection.   The QAOA is a gate-model ansatz designed to produce a quantum state that minimizes (or maximizes) the expectation value of a classical cost function.  Using the notation from Eq.~\eqref{eq:HTIM}, the level $p$ QAOA produces an approximation $C^*$ to the optimal value of the classical cost function,
\begin{align}
U(\beta, \gamma) &= e^{i \beta H_X} e^{i \gamma H_Z}\\
|\psi(\boldsymbol{\gamma},\boldsymbol{\beta}) &= \left(\prod_{k =1}^p U(\beta_k , \gamma_k) \right) |+^n\rangle\\
C^* = &\min_{\boldsymbol{\gamma},\boldsymbol{\beta}} \; \; \langle\psi(\boldsymbol{\gamma},\boldsymbol{\beta}) | H_Z | \psi(\boldsymbol{\gamma},\boldsymbol{\beta}) \rangle
\end{align}
where $\boldsymbol{\gamma} = (\gamma_1 , ... , \gamma_p) ,\boldsymbol{\beta} = (\beta_1,...,\beta_p)$ are the angles that parameterize the circuit.  Various heuristic methods for choosing these angles have been considered, and for small values of $p = O(1)$ the optimization can be done exactly~\cite{szegedy2019qaoa}.  
First, we note that both coherent \emph{adiabatic} TH-HI and QAOA are able to obtain limited quantum speedups against classical simulated annealing (SA) in toy problems~\cite{bapat2018bang}. However, this speedup relative to SA alone is of limited value. Indeed, additional toy examples demonstrate the ability of QAOA to outperform coherent adiabatic TF-HI as well as SA~\cite{streif2019comparison}, though this work did not consider diabatic TF-HI. For the latter, we already noted in Sec.~\ref{CHTF-HI} that coherent diabatic TF-HI can solve versions of the spike problem in time $\mathcal{O}(1)$~\cite{Muthukrishnan:2015ff}. The same holds for QAOA~\cite{bapat2018bang}. 

There exist negative results about QAOA: it does not in general outperform the (classical) Goemann-Williamson algorithm for certain instances of the MaxCut problem at any finite depth~\cite{bravyi2019obstacles}, and $p = 1$ is outperformed by local classical algorithms~\cite{hastings2019classical}.  QAOA suffers reachability deficits (the inevitable need for large depth) in cases with a high clause-to-variable ratio~\cite{akshay2019reachability}.  

A very important result from our perspective is that despite results on the optimality of bang-bang control~\cite{Yang:2017aa}, it appears that the optimal QAOA angle parameters digitize an asymptotically smooth curve; see Fig.~\ref{fig:QAOA}. QAOA is closely related to an optimized diabatic QA path, with an explicit correspondence given in Ref.~\cite{zhou2018quantum} between QAOA angle parameters and a TF-HI annealing schedule $H(t) = (1-f(t))H_Z + f(t) H_X$ via:
\bes
\begin{align}
T &= \sum_{k = 1}^p (|\gamma_k| + |\beta_k|)\\
t_i &= \sum_{k = 1}^i |\gamma_k| + |\beta_k| - \frac{1}{2}(|\gamma_i| + |\beta_i|)\\
&f\left(t_i\right)= \frac{\gamma_i}{|\gamma_i| + |\beta_i|}
\end{align}
\ees
An explanation for this apparent coincidence was reached in~\cite{brady2020optimal}, who used optimal control theory to show that generically, given a fixed amount of time, the optimal procedure has the pulsed (or ``bang-bang") structure of QAOA at the beginning and end but can have a smooth, adiabatic annealing structure in between. Through simulations of various transverse field Ising models, they demonstrated that bang-anneal-bang protocols are more commonly optimal than either pure QAOA or pure adiabatic protocols. An example is shown in Fig.~\ref{fig:bang-anneal-bang}. However, finding the optimal schedule remains a hard problem. Note that after initializing the system in the ground state of the initial Hamiltonian, the schedule starts with a constant \emph{final} Hamiltonian, and ends with a constant \emph{initial} Hamiltonian, with the goal of driving the system to the ground state of the final Hamiltonian. It is possible that the non-zero length of the initial and final constant segments is a finite size effect. It is certainly expected, by the adiabatic theorem, that these segments will shrink to zero in the limit of large $T$ and $p$ (QAOA circuit depth).

The overall conclusion is currently that QAOA and coherent diabatic TF-HI are comparable with regard to positive and negative evidence for a limited quantum speedup relative to specific classical algorithms, and multiple results (with accompanying mathematical explanations) appear to indicate that \emph{optimized intermediate-depth QAOA schedules are converging to a Trotterized version of a continuous curve, i.e., the angles are Trotterizing a continuous-time transverse-field Ising evolution that corresponds to slightly non-monotonic versions of diabatic TF-HI.} This means that existing evidence is pointing toward QAOA achieving optimal performance only in the limit in which it becomes equivalent to coherent diabatic TF-HI, with the possible exception of the initial and final segments of the anneal. These results appear to remain valid even for weakly-decoherent systems~\cite{DCL:inprep}. Finding the optimal angles for QAOA then becomes equivalent to finding the optimal schedule for coherent diabatic TF-HI.

\section{The Role of Nonstoquasticity in Classical Intractability}
\label{sec:nonstoq}

Recall that a Hamiltonian is called stoquastic if there is a choice of a local basis in which the Hamiltonian matrix elements are real and nonpositive~\cite{Bravyi:QIC08}.   Otherwise the Hamiltonian is called nonstoquastic, and the inevitable positive or complex off-diagonal matrix elements of the Hamiltonian lead to the QMC sign problem.  Quantum Monte Carlo methods enable the estimation of local observables for thermal equilibrium states of a stoquastic Hamiltonian using a relatively small number of samples from a probability distribution over paths of basis states.\footnote{Even in this case the time needed to obtain each sample is related to the equilibration of a Markov chain, and this equilibration tends to dominate the runtime in the simulation of transverse-Ising spin glasses.}  
The sign problem transforms this probability distribution over paths into a pseudo-probability distribution, i.e., one which includes negative or complex ``probabilities,'' and in this case it is no longer efficient to estimate observables using a small number of samples due to cancellations.  

Even in cases for which the Hamiltonian is stoquastic, but it is presented in a form in which this stoquasticity is unapparent, it can be NP-hard to find the basis that ``cures the sign problem'' by making all of the Hamiltonian matrix elements real and nonpositive~\cite{Marvian:2019aa,klassen2019hardness}.  Performing QMC in any other basis will generically create a sign problem, and in fact measuring equilibrium states of stoquastic Hamiltonians in a rotated basis can sample distributions that are classically intractable~\cite{fujii2018quantum}.  While there is a close qualitative link between positive Hamiltonian matrix elements and a sign problem, the quantitative severity of the sign problem is in general difficult to estimate from the form of the matrix elements.  Therefore a quantitative examination of this route to classical intractability should examine the statistical severity of the sign problem~\cite{Gupta_2020}.  The definition of nonstoquasticity was motivated by complexity theoretic considerations, while the sign problem is the more directly meaningful measure of the hardness of QMC simulations.

A related notion is that of developing qubits that can statically emulate a vector spin-$1/2$ system, with the ability to independently tune dipole-dipole interactions for the $X,Y$ and $Z$ components~\cite{Kerman:2019aa}.  The ability to control arbitrary $2$-local interactions would enable the emulation of nonstoquastic Hamiltonians as well as stoquastic Hamiltonians of a more general form than transverse-field Ising models.  This latter capability is the one needed for Hamiltonian error suppression using stabilizer subsystem codes~\cite{Jiang:2015kx,Marvian-Lidar:16,marvian2019robust}, since to achieve universality they require both $\pm XX$ and $\pm ZZ$ interactions (though the penalty terms for the only fully two-local Hamiltonian error suppression protocol~\cite{marvian2019robust} are stoquastic).  These error suppression protocols implemented in the setting of universal adiabatic quantum computing are arguably the most compelling reason to pursue qubit technologies that enable static dipole-dipole interactions along multiple vector components. 

An argument that is made sometimes to motivate vector interactions is to realize strong multi-spin fluctuations (of either sign), even though in equilibrium this does not have a provable advantage in terms of classical simulatability. The argument is that one should explore the potential power of multi-spin fluctuations in heuristic applications out of equilibrium. However, a serious concern about this argument is that multi-spin fluctuations will only confer a similar advantage as adding multi-spin flips to a classical stochastic process like simulated annealing (i.e., instead of flipping one bit at a time, one proposes to flip two or more bits at a time). This could make a difference at small system sizes, but will stop making much of a difference at the intermediate scale.  

Returning to the role of nonstoquastic Hamiltonians in quantum enhancement, it is strongly believed that computational basis measurements of thermal states and ground states of nonstoquastic Hamiltonians give rise to classically intractable distributions, but the prospects for algorithmic enhancement are less well understood.  For ground states, formal evidence of this intractability comes from the fact that various families of $2$-local nonstoquastic Hamiltonians are universal for adiabatic computation and also have ground state energies that are QMA-hard to approximate.\footnote{QMA stands for ``quantum Merlin-Arthur", the natural quantum generalization of the classical complexity classes NP and MA. Informally, QMA is the class of problems that can be efficiently checked on a quantum computer given a ``witness" quantum state related to the answer to the problem~\cite{Kitaev:book}.}  However, these formal results require substantial overhead in the form of additional qubits and perturbative gadgets that require precise control of the qubit couplings across orders of magnitude.%
\footnote{More specifically, there is a large overhead in the number of qubits needed to represent clocks and ancillas for gadgets~\cite{Biamonte:07}, and there are unrealistic variations in coupling strength needed for perturbative gadgets due to the use of first order perturbation theory (e.g., a 10 to 1 ratio of couplings)~\cite{Lloyd:2016}.}
Therefore, the general belief in the classical intractability of sampling nonstoquastic equilibrium states rests on practical evidence, that is the QMC sign problem and the lack of other candidate algorithms to efficiently sample from these distributions.

There are some specific cases in which nonstoquastic Hamiltonians can improve ground state adiabatic optimization by turning a phase transition from first to second order (exponentially small to polynomially small gap, respectively)~\cite{susa2017relation}, and some understanding of this has been developed in terms of ground states in symmetric and antisymmetric subspaces~\cite{Albash:2019aa}.  However, it is known that multi-modal ground state distributions inevitably lead to small gaps and this problem cannot be alleviated by nonstoquastic Hamiltonians~\cite{Crosson:2017aa}.  More recently, techniques in random matrix theory were used to show that stoquastic Hamiltonians have much larger spectral gaps between the ground state and first excited state with high probability, and this effect is also confirmed numerically for local Hamiltonians~\cite{crosson2020designing}.  This latter study is based on the notion that any nonstoquastic Hamiltonian can be ``de-signed'' (i.e., have its positive signs removed) into a corresponding stoquastic Hamiltonian that has a larger spectral gap, with high probability.  It was also noted numerically that these de-signed Hamiltonians have a shorter time-to-solution even when the Hamiltonian interpolation proceeds diabatically.  The work on de-signed Hamiltonians indicates that $-XX$ (stoquastic) interactions are superior to $+XX$ (nonstoquastic) interactions with high probability. It may also be the case more generally than~\cite{susa2017relation} that $-XX$ interactions lead to improvements over the TF-HI form.  This supports the notion that vector dipole-dipole interactions (needed to generate strong $\pm XX$ interactions of either sign) can improve diabatic Hamiltonian interpolation, apart from whether these interactions are used to generate a nonstoquastic Hamiltonian.

In summary, \emph{nonstoquastic Hamiltonians create a QMC sign problem and give rise to classically intractable measurement distributions, but out-of-equilibrium DQA dynamics also achieve this with comparable confidence.}  Quite separately from the question of nonstoquasticity, we currently have two reasons to support the development of vector dipole-dipole interactions, these being universal Hamiltonian computation with error suppression, and evidence that more general $2$-local stoquastic Hamiltonians improve diabatic Hamiltonian interpolation. Neither of these potential avenues to enhancement rely specifically on nonstoquasticity ($+XX$ as opposed to $-XX$ interactions).

\section{Sampling Applications and Machine Learning}
\label{sec:formal}

\subsection{Classical Intractability of Sampling}
\label{sec:intrac}
In this section we wish to explain why despite the perceived virtues of quantum supremacy associated with the gate-model~\cite{Arute:2019aa}, the lack of known supremacy results for the DQA model should not be viewed as a negative for the sampling algorithms we consider.

A key point we have already made is that classical intractability of a quantum process does not imply that it can provide enhancement, and this point is particularly salient in the context of quantum supremacy, as we now explain.  All of the existing arguments for the classical intractability of quantum sampling problems take the following form. 
\begin{enumerate}
\item A certain quantum process, if supplemented with the unrealistic capability of postselecting on exponentially small amplitudes, would be universal for postselected quantum computation.  
\item If the original quantum process could be efficiently classically simulated with sufficient precision, then the version of the process with postselection could be efficiently simulated by postselected classical computation.  
\item	Therefore, efficiently simulating the quantum process classically would imply equality of postselected classical computation (which is contained in the 3rd level of the polynomial hierarchy) and postselected quantum computation (which is outside this hierarchy) and this is nearly as implausible as P = NP. 
\end{enumerate}
There are two main weaknesses with this line of argument, both of which are widely acknowledged in the formal works that treat these results~\cite{Harrow:2017aa}. The first weakness is the level of precision required for item 2. To apply these arguments without additional assumptions it is required that the classical algorithm approximately samples the output of the quantum process with \emph{exponentially} small error (in the trace norm). This level of precision is clearly unrealistic, but it also represents a worst-case analysis.  Therefore a later result~\cite{Bremner:2016aa} uses additional assumptions 
about the ensemble of distributions resulting from a class of quantum processes to perform worst-case to average-case reductions~\cite{Bouland:2018aa,Movassagh:2019aa}, to argue that classically sampling from a distribution that is within \emph{constant} trace-norm error from the output of the quantum process also would imply collapse of the polynomial hierarchy and is therefore unreasonable.

While we regard these additional assumptions as plausible, it is substantially more difficult to build device architectures whose quantum processes are manifestly compatible with them.  For example, there are now multiple works demonstrating sampling-type efficient classical simulations for 2D constant-depth quantum circuits~\cite{bravyi2019classical,napp2019efficient} (constant depth implies low entanglement width across the circuit), while for unrestricted connectivities we still expect the arguments above to apply.  Notable attempts at achieving sampling supremacy in a 2D architecture based on short-time Hamiltonian dynamics have been made~\cite{haferkamp2019closing,fujii2018quantum}, but these require rotated basis measurements and have more in common with measurement-based quantum computation. \emph{The main point of this discussion is that one must reject claims of formal evidence for classical intractability of a device process unless all of the required formal assumptions are verified, and the violation of any of them nullifies the convincing power of these arguments.}%
\footnote{Anti-concentration is a key assumption in some supremacy results (e.g., in boson sampling~\cite{Aaronson_2011}) needed to go from hardness of approximating output amplitudes to sampling, but it has been proved for IQP circuits and random circuit sampling~\cite{Hangleiter2018anticoncentration}. It is possible that other assumptions will similarly be removed as the field progresses.}

The second weakness of these hardness of sampling arguments is that they are necessary but not sufficient to imply enhancement.   This was very clear in the case of the early quantum supremacy proposals based on IQP circuits~\cite{bremner} and linear optics~\cite{Aaronson_2011}, but with the rise of NISQ era gate-model devices this point has becoming increasingly clouded by the rush to claim a quantum advantage, particularly in the application areas of optimization and machine learning.
For example, there is a tendency to equate formally-supported classical intractability of sampling these distributions with the idea that these distributions are “computationally powerful.”    But a different perspective is that these distributions are only hard to sample because we lack accessible classical descriptions of them. Gibbs distributions of classical spin glasses are hard to sample from even though we can write down a succinct expression for the probability $p$ of every spin configuration $x$,
$p(x) = \frac{e^{-E(x)}}{Z}$.
Although the partition function $Z$ is computationally intractable, knowledge of the unnormalized probability density $e^{-E(x)}$ is the basis for applying state-of-the-art Markov-chain Monte-Carlo (MCMC) methods like parallel tempering with iso-energetic cluster moves (PT-ICM)~\cite{Houdayer:2001aa,PhysRevLett.115.077201}.  

This raises the question, are output distributions of these quantum devices hard to sample because they are just fundamentally difficult distributions to sample from (like spin glass Gibbs states), or is it only the lack of a concise description of the unnormalized density that prevent classical simulation?   To phrase this question mathematically, suppose we had an efficiently computable function $w$ such that $\pi(x) = w(x) / A$
for the quantum device output distribution $\pi$.  If we had access to such a $w$ then we could apply classical MCMC algorithms and ask whether they converge efficiently to $\pi$.   In the case of constant depth circuits the answer to this is already known: a Metropolis algorithm with access to the function $w$ above would always converge in $\mathcal{O}(n \log n)$ steps because of the vertex expansion properties of the output distributions of constant-depth circuits~\cite{eldar2017local,Crosson:2017aa}.  In brief, these constant depth output distributions must be either single-peaked or highly delocalized, and the existence of multiple peaks
far apart in Hamming distance, which would be necessary to foil the Metropolis algorithm, is ruled out for this class of circuits. 

This argument proves that the classical intractability of output distributions of constant-depth circuits is only due to our lack of a classical description of the distribution, and not the nature of the distribution itself (since once given the density $w$ in an efficiently computable form, we could sample the distribution classically).  Another potential example of this could be the ``speckle-like" output of random quantum circuits~\cite{Arute:2019aa}.  These distributions do not have the kind of far-separated peaks (which in the Gibbs state manner of thinking would correspond to deceptive local minima of a cost function) that we know from experience are hard to classically sample with MCMC.  Again it may be the lack of a classical description, rather than the nature of the distribution itself, that is causing the classical intractability.

These points are counter to the general exuberance about sampling applications that we observe in the NISQ era.  There may be a coming realization
in the gate-model community 
that pinning one's hopes on 
``quantumness" is not enough, and that one must dive much deeper into the nature of these output distributions to see an indication of whether they are useful.   The QA community has already reached this stage years ago, and this is directly related to the large body of negative results that has been accumulated, alluded to in the Introduction.  The time-lag between different device technologies is currently causing their utility to be judged by somewhat different standards, and in the pre-fault-tolerant era, any optimism we maintain for one platform should be maintained for the other.    

\subsection{Machine Learning}

Quantum machine learning (QML) is a good example of an area where initial exuberance surrounding gate-model algorithms gave way to a more realistic assessment, after several dequantisation results were found wherein classical algorithms performed as well as QML~\cite{tang2018quantuminspired,Tang_2019,gilyn2018quantuminspired}. Quantum annealing has not fared any better in this regard, but a common thread in all of QML is that there are strong reasons to believe that it is classically hard to simulate the training process, since it relies on quantum dynamics. For this to apply to transverse-field Ising models, it is again essential to invoke DQA.

\subsubsection{Training classifiers using quantum annealing}
A binary classifier can be thought of as a function that returns a binary value given an input from a dataset, which it attempts to sort into two classes. The classifier is a weighted linear combination of ``features" (also known as weak classifiers), functions defined over the same dataset that are by design sensitive only to partial properties of the data (e.g., in image recognition the features could be functions that respond to the presence of a certain color).  The problem is then to find the optimal values of the feature weights, so that the classifier achieves high accuracy over a labeled training dataset. Its performance is subsequently evaluated over a test dataset, returning a label for each element. The training stage of this classical ``boosting" algorithm~\cite{Freund:1997aa} can be replaced by runs of a quantum annealer, via an appropriate relaxation of the optimization problem in terms of quadratic cost function~\cite{Neven1,Pudenz:2013kx}. Each state returned by the quantum annealer is then a set of feature weights (e.g., one weight per spin state), which can be averaged to generate the eventual classifier. This includes excited states observed at the end of each annealing run, and such states arise from either diabatic or thermal transitions. With accuracy as the performance metric rather than time-to-solution, this approach was tried on datasets ranging from simulated Higgs-boson events~\cite{Mott:2017aa,Zlokapa:2019aa} to transcription factor binding to DNA~\cite{Li:comp-bio-2017,Willsch:2020aa}, and genomic cancer data~\cite{Li:2019aa,Jain:2020aa}. No advantage was observed for the quantum annealer, but accuracy was comparable to state of the art classical machine learning methods for the smallest training datasets. It remains to be seen whether this result will translate to a future advantage for QA-based classifier training; given our discussion so far, it appears that the only viable path to such a result would be through the role played by excited states in constructing the classifiers.

\subsubsection{Quantum Boltzmann Machines.}
Quantum Boltzmann Machines (QBMs) are based on sampling thermal equilibrium states of quantum Hamiltonians~\cite{Amin:2016}, and as such can be viewed as belonging to the category of weakly-decoherent adiabatic models.  The most well-studied QBMs are based on transverse-field Ising Hamiltonians, but classes of Hamiltonians with more general (non-stoquastic) off-diagonal terms have also been considered. The results on QBMs are largely heuristic and small-scale numerical, and there is currently no evidence pointing at a quantum advantage. However, it has been shown that classical computers cannot simulate the training process of a QBM in general unless BQP=BPP~\cite{Kieferova:2017aa}.

\subsubsection{DQA Ising Born Machines.}
Stochastic neural networks formed from measurement distributions of non-thermal quantum states are referred to as Born machines~\cite{Liu:2018aa,e20080583,Benedetti:2019aa}, after the Born rule which converts quantum amplitudes into probabilities.  Quantum Circuit Ising Born Machines (QCIBMs) are generative learning models based on parameterized quantum circuits consisting of gates that are generated by transverse-field Ising interactions~\cite{coyle2019born}. The unsupervised learning task they solve is to generalize from a finite set of samples drawn from a data set, by learning their underlying probability distribution.
The parameterized circuits proposed for QCIBMs involve the same kinds of alternating sequences of mixing operators and phase separation operators that are used in QAOA~\cite{Farhi:2014aa,a12020034}, with the main difference being the cost function that is used to variationally tune the parameters of the quantum circuit.   In QAOA this cost function is a classical combinatorial optimization problem, while for the IBMs the cost function is taken to be one of various tractable notions of distance between the QCIBM output distribution and the target distribution which one intends to learn.  

A DQA-IBM would simply replace the parameterized quantum circuit with a smooth annealing schedule depending on some finite number of parameters (e.g., some number of Fourier components of the coupling functions in the annealing schedule).  All of the training methods proposed in~\cite{coyle2019born} for QCIBMs are based on computational basis measurements and classical post-processing to tune the parameters in the circuit, and these methods could be equally well applied to training the parameters of an annealing schedule.

It was shown in~\cite{coyle2019born} that QCIBMs cannot, in the worst case, and up to suitable notions of error, be simulated efficiently by a classical device. The same would be true of DQA-IBMs, thus providing a path to a quantum advantage in training IBMs. Whether this would lead to the ability to learn distributions more efficiently than any classical algorithm (``quantum learning supremacy") remains an open question.

\section{Reducing Noise and Errors}
\label{sec:noise}

\subsection{Long-term Challenges for Gate-Model Optimization}

An oft-cited reason for investing in the gate-model instead of QA is that the former supports a well-developed theory of fault-tolerance.  Even if a full theory of fault-tolerance is developed for continuous-time Hamiltonian evolutions, we are most likely to build eventual fault-tolerant quantum computers (FT-QCs) in whichever is most efficient in terms of fault-tolerant overhead.  There is no doubt that fast operations and intermediate measurements are advantageous for the localization and correction of errors in quantum systems, and so the gate-model is likely to maintain an advantage in this area in any case.  When it comes to exponential speedups, especially in the application area of quantum simulation, there is a nearly unanimous consensus that fault-tolerant gate-model quantum computers will one day achieve a useful quantum enhancement.

However, when we turn to optimization and machine learning there is less reason to be confident that FT-QCs will eventually achieve useful enhancement.  In the worst-case of NP-hard problems it is expected that quantum computers can provide at most a polynomial speedup, and assuming the Quantum version of the Strong Exponential Time Hypothesis~\cite{aaronson2019quantum} we expect that 3-SAT with $n$ variables inevitably takes time $\Omega(2^{n/2})$.  While such a quadratic speedup is clearly useful asymptotically, it needs to be closely compared with fault-tolerant overhead at the finite-system sizes of applications.  

A recent review on the prospects for rigorous speedups in optimization with FT-QC~\cite{campbell2019applyingquantum} at finite system sizes reports results that can be seen as discouraging (see also Ref.~\cite{s2020compilation}).  One of the few general-purpose rigorous quantum algorithms for optimization in the gate-model is Grover's search, which treats solutions as marked elements.  This could be applied to either exact optimization (e.g., mark the satisfying assignments of a k-SAT instance) or approximate optimization (e.g., mark all assignments that violate fewer than $m$ clauses), and in general yields a quadratic speedup.  The other more sophisticated class of algorithms considered in \cite{campbell2019applyingquantum} are based on backtracking algorithms, which are exact optimization algorithms that can be regarded as exploring a tree of partial solutions to an optimization problem and pruning branches that violate constraints.  These quantum backtracking algorithms obtain a quadratic speedup in terms of the number of nodes in the partial solution tree, which generally remains exponential but typically has size less than $2^n$. Arguments are given for constant factor speedups, but the number of physical qubits used is extremely large, and improved fault-tolerance methods will likely be needed to make these results practical. In particular, the quantum advantage disappears if one includes the cost of the classical processing power required to perform decoding of the surface code using current techniques.

\subsection{Hamiltonian Error Suppression}
\label{sec:HES}

The standard gate-model approach to fault-tolerance requires families of codes with asymptotically growing size and distance, fast measurements used to detect errors, and a large classical processing overhead to decode and correct errors between each quantum clock cycle.  These capabilities enable arbitrarily long computations if the errors are below some fixed threshold, but depending on the finite size of applications and the expected speedup the overheads required may be prohibitive. 
Hamiltonian Error Suppression (HES)~\cite{Bacon:01,jordan2006error,Young:2013fk,PAL:13,PAL:14,MNAL:15,Bookatz:2014uq,Jiang:2015kx,Vinci:2015jt,vinci2015nested,Mishra:2015,Marvian-Lidar:16,Marvian:2016aa,Matsuura:2016aa,Marvian:2017aa,Matsuura:2018,Lidar:2019ab,marvian2019robust,Pearson:2019aa} is 
an alternative to the standard approach that,
in contrast, uses error \emph{detecting} codes with a fixed or growing~\cite{vinci2015nested} distance, and does not require any intermediate measurements or classical processing.  HES is designed for continuous-time Hamiltonian computation and enforces containment in a logical code space using energy penalties that suppress transitions to states outside of the code space. Note that HES is also the natural framework for preserving $\mc{C}$-coherence (Def.~\ref{def:C-coherence}), since the logical code space can be a degenerate ground subspace but also a non-degenerate subspace with a finite energy width.

The strongest theoretical results on HES to date establish that using an energy-penalty strength that grows only logarithmically in system size, at a fixed temperature, 
errors arising from coupling to a Markovian environment can be exponentially suppressed in the penalty strength for arbitrary long times, as long as the gap closes no faster than inverse polynomial in the system size~\cite{Lidar:2019ab}. This, however, requires four-local interactions for universality. A fully two-local scheme has also been developed, at the expense of giving up exponential error suppression~\cite{marvian2019robust}.
 
Major open questions include whether these results can be extended to gaps closing exponentially, whether fully two-local schemes can provide exponential error suppression,
and whether theoretical bounds beyond quadratic scaling of error cancellation with the number of physical qubits per logical qubit~\cite{Young:2013fk} can be established for the degree of protection HES provides against intrinsic analog control errors. While promising empirical results were reported along these lines~\cite{Pearson:2019aa,vinci2015nested}, such a result would be essential in order to approach a semblance of fault tolerance in HES.  Hardware implementations of HES would enable the investigation of these questions beyond the range of classical simulation.  
Most pressing in this regard would be the addition of a constant-in-time $-XX$ coupling to the Hamiltonians in Eq.~\eqref{eq:h}, since this would suffice in order to achieve fully two-local HES~\cite{marvian2019robust}.  \emph{Hamiltonian error suppression is one of the clearest examples of a moonshot in modern quantum information science: if it works well (better than predicted by worst-case theoretical bounds) then it could dramatically reduce the expected overhead needed for implementing quantum algorithms at the application scale.}

Other alternative schemes to investigate include hybridizing continuous-time Hamiltonian dynamics with some fast measurement capabilities.  Measurements are useful because they can directly remove entropy from the quantum system.   This can also be done by continually resetting ancilla qubits to the state $|0\rangle$ and interacting them with the system coherently.  A promising recent approach combines HES with weak measurement of the code Hamiltonian~\cite{atalaya2019error,atalaya2020continuous}.

\section{Summary}
\label{sec:summary}

To guide future research into the power of quantum annealing, broadly defined here in terms of continuous-time evolution under the transverse-field Ising model, including various generalizations such as reverse annealing, we asked under which conditions efficient competing classical algorithms already exist or cannot be expected to be discovered. 

To answer this question we distinguished four cases: (i) coherent-adiabatic, (ii) weakly decoherent-adiabatic, (iii) weakly decoherent diabatic, and (iv) coherent and $\mc{C}$-coherent diabatic (Def.~\ref{def:C-coherence}). We also distinguished between forward and reverse annealing protocols in these four cases. In the forward QA case, we argued that the most promising of the four is the coherent and $\mc{C}$-coherent diabatic case. It is in this case that there is no known classical algorithm that can reasonably be expected to compete with a quantum annealing device, and the same cannot be said of cases (i)-(iii). The evidence in the reverse-QA case is more encouraging, and we concluded that only the weakly-decoherent adiabatic case is unpromising (see Fig.~\ref{fig:4-cases} for a summary).
 Of course, the fully coherent diabatic case is an idealization. We expect it to grow in importance in providing a setting for proving theoretical quantum advantage results. The $\mc{C}$-coherent diabatic case is more practically useful, but will require Hamiltonian error suppression in order to be enforced in physical systems. The associated effort will be significantly less than what is required to achieve fault tolerance in the gate model.

Very recent theoretical (both analytical and numerical) results provide additional reasons to believe that the coherent/$\mc{C}$-coherent-diabatic case is optimal. These recent studies~\cite{zhou2018quantum,pagano2019quantum,brady2020optimal} make the case that QAOA converges to a diabatic QA protocol, in the sense that the optimal choice of QAOA angles approximates a parametrization of a continuous annealing schedule. 

It is important to emphasize that the diabatic case does not require pulsed interactions (unlike QAOA), i.e., the annealing schedules ${A(t),B(t)}$ can be slowly varying compared to the gate-model, since this should still be sufficient to compete with QAOA. Rather, diabaticity is a consequence of a violation of the adiabatic condition in the sense that the total evolution time $T$ is short on the timescale set by energy gaps encountered along the annealing path. 

We also addressed the question of the need for nonstoquastic interactions. On the one hand it is known that 
nonstoquastic Hamiltonians create a QMC sign problem and give rise to classically intractable measurement distributions. On the other hand, stoquastic out-of-equilibrium DQA also achieves this with comparable confidence. In this regard there is no clear advantage to nonstoquastic interactions. The main theoretical reason to support the development of nonstoquastic Hamiltonians is that they enable relatively realistic architectures for universal adiabatic or Hamiltonian computation with error suppression. There is also some evidence that nonstoquastic Hamiltonians can remove first order quantum phase transitions, but this can also be accomplished in the stoquastic setting using, e.g., reverse annealing. The latter is a promising heuristic when used as an iterative algorithm that is allowed to improve from one iteration to the next by starting a new anneal cycle from an excited state arrived at diabatically or thermally in the previous cycle.

Sampling applications, in particular various forms of machine learning such as Ising Born machines, for which classical hardness results can be proven, also benefit from being cast in the DQA setting.

In conclusion, we advocate for a concerted theoretical and experimental effort focused on diabatic quantum annealing protected via Hamiltonian energy suppression as a fast path towards quantum advantage that is less resource intensive than the gate model and more promising that the adiabatic model.

\acknowledgments
We are grateful to many of our 
colleagues in the IARPA-QEO and DARPA-QAFS programs, in particular A. Kerman, E. Rieffel, F. Wilhelm, and K. Zick, for their comments and insights. We also acknowledge helpful discussions with Dr. Richard Harris about D-Wave's reverse annealing protocol.
The research is based upon work (partially) supported by the Office of
the Director of National Intelligence (ODNI), Intelligence Advanced
Research Projects Activity (IARPA) and the Defense Advanced Research Projects Agency (DARPA), via the U.S. Army Research Office
contract W911NF-17-C-0050. The views and conclusions contained herein are
those of the authors and should not be interpreted as necessarily
representing the official policies or endorsements, either expressed or
implied, of the ODNI, IARPA, DARPA, or the U.S. Government. The U.S. Government
is authorized to reproduce and distribute reprints for Governmental
purposes notwithstanding any copyright annotation thereon.


%

\end{document}